\documentclass[journal=esthag,manuscript=article]{achemso}
\setkeys{acs}{email=true,keywords=true}

\usepackage[version=3]{mhchem} %

\usepackage{xspace}
\usepackage{ulem}
\usepackage{multirow}
\makeatletter
\DeclareRobustCommand\onedot{\futurelet\@let@token\@onedot}
\def\@onedot{\ifx\@let@token.\else.\null\fi\xspace}

\def\eg{\textit{e.g}\onedot} 
\def\ie{\textit{i.e}\onedot}

\def\etal{\textit{et al}\onedot}
\makeatother

\usepackage{xcolor}
\usepackage[linesnumbered,ruled,vlined]{algorithm2e}
 \usepackage{stmaryrd}
\usepackage{nameref}
\usepackage{subcaption}
\newbox{\bigpicturebox}

\newcommand{\nada}[1]{}

\usepackage{amsmath}
\DeclareMathOperator*{\argmin}{arg\,min}

\usepackage{graphicx}
\author{Thibaud Ehret}
\email{thibaud.ehret@ens-paris-saclay.fr}
\affiliation[CB]
{Université Paris-Saclay, CNRS, ENS Paris-Saclay, Centre Borelli, Gif-sur-Yvette, 91190, France}

\author{Aurélien De Truchis}
\affiliation[K]
{Kayrros SAS, Paris, 75009, France}

\author{Matthieu Mazzolini}
\affiliation[K]
{Kayrros SAS, Paris, 75009, France}

\author{Jean-Michel Morel}
\affiliation[CB]
{Université Paris-Saclay, CNRS, ENS Paris-Saclay, Centre Borelli, Gif-sur-Yvette, 91190, France}

\author{Alexandre d'Aspremont}
\affiliation[CNRS]
{CNRS, Ecole Normale Sup\'erieure, Paris, 75230, France}
\alsoaffiliation[K]
{Kayrros SAS, Paris, 75009, France}

\author{Thomas Lauvaux}
\affiliation[LSCE]
{Laboratoire des Sciences du Climat et de l’Environnement, CEA, CNRS, UVSQ/IPSL, Saint-Aubin, 91190, France}

\author{Riley Duren}
\affiliation[UA]
{Arizona Institutes for Resilience, University of Arizona, Tucson, AZ, 85721, USA}
\alsoaffiliation[CM]
{Carbon Mapper, Pasadena, CA, 91105, USA}

\author{Daniel Cusworth}
\affiliation[UA]
{Arizona Institutes for Resilience, University of Arizona, Tucson, AZ, 85721, USA}
\alsoaffiliation[CM]
{Carbon Mapper, Pasadena, CA, 91105, USA}

\author{Gabriele Facciolo}
\affiliation[CB]
{Université Paris-Saclay, CNRS, ENS Paris-Saclay, Centre Borelli, Gif-sur-Yvette, 91190, France}

\title{Global Tracking and Quantification of Oil and Gas Methane Emissions from Recurrent Sentinel-2 Imagery}

\abbreviations{IR,NMR,UV}

\begin{document}

\begin{tocentry}
    \includegraphics[width=\linewidth]{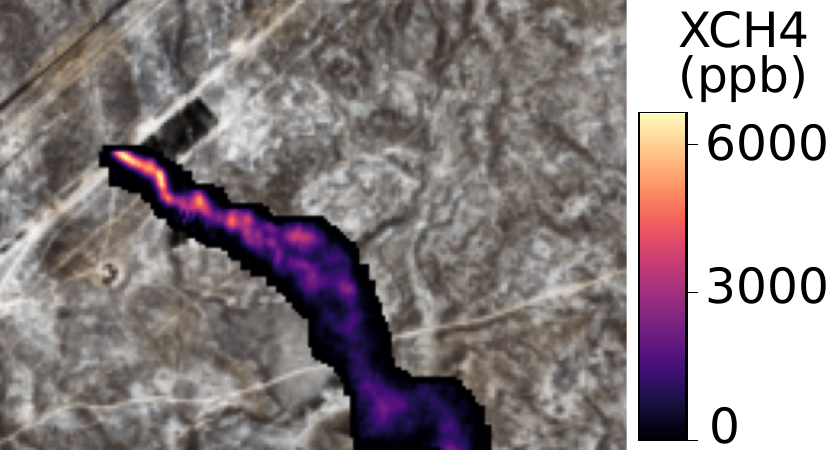}
\end{tocentry}

\begin{abstract}
Methane ($\text{CH}_4$) emissions estimates from top-down studies over oil and gas basins have revealed systematic under-estimation of $\text{CH}_4$ emissions in current national inventories. Sparse but extremely large amounts of $\text{CH}_4$ from oil and gas production activities have been detected across the globe, resulting in a significant increase of the overall O\&G contribution. However, attribution to specific facilities remains a major challenge unless high-spatial-resolution images provide the sufficient granularity within O\&G basin.
In this paper, we monitor known oil-and-gas infrastructures across the globe using recurrent \emph{Sentinel-2} imagery to detect and quantify more than 1200 $\text{CH}_4$ emissions. In combination with emissions estimates from airborne and \emph{Sentinel-5P} measurements, we demonstrate the robustness of the fit to a power law from 0.1 t$_{\text{CH}_4}$/hr to 600 t$_{\text{CH}_4}$/hr. 
We conclude here that the prevalence of ultra-emitters ($>$ 25t$_{\text{CH}_4}$/hr) detected globally by \emph{Sentinel-5P} directly relates to emission occurrences below its detection threshold in the range $>$ 2t$_{\text{CH}_4}$/hr, which correspond to large-emitters covered by \emph{Sentinel-2}. We also verified that this relation is also valid at a more local scale for two specific countries, namely Algeria and Turkmenistan, and the Permian basin in the US.
\end{abstract}

\section{Synopsis}

This work shows the global methane emission monitoring potential of Sentinel-2 and compare to those of Sentinel-5 and local airborne studies.

\section{Introduction}

The detection of large and frequent methane ($\text{CH}_4$) emissions linked to oil and gas production has raised concerns in the ability of natural gas to effectively reduce greenhouse gas (GHG) emissions as a substitute to coal~\cite{brandt2014methane,duren2019california, mckeever2019detection,varon2019satellite,zhang2020quantifying,guha2020assessment,irakulis2021satellite,cusworth2021intermittency}. Over a 20-year horizon, a $\text{CH}_4$ molecule has a global warming potential close to 80 times larger than carbon dioxide (CO$_2$)~\cite{IntergovernmentalPanelonClimateChange2021}. A large part of the $\text{CH}_4$ emissions could be controlled or avoided, as they come primarily from maintenance operations at oil rigs, pipelines, or well pads, and from equipment failures~\cite{lyon2016aerial}. 

In order to detect and quantify GHG fossil fuel emissions produced by human activities, several satellites have been placed in orbit over the past ten years (e.g. GOSAT, OCO-2, TROPOMI), allowing a persistent monitoring of carbon dioxide and methane abundance.
The \emph{Sentinel-5P} (TROPOMI) satellite mission~\cite{veefkind2012tropomi} provides hyper-spectral images in the short-wave infrared (SWIR) spectrum in which $\text{CH}_4$ is a strong absorber. It provides daily $\text{CH}_4$  column mole fractions over the whole globe at relatively low spatial resolutions (5-7 km) revealing multiple individual cases of very large emissions (e.g. Pandey~\etal~\cite{pandey2019satellite}) and regional basin-wide anomalies~\cite{schneising2020remote, barre2021systematic}. However, due to its relatively low spatial resolution%
, this mission remains inadequate to observe small emissions ($<$ 25 t$_{\text{CH}_4}$/hr) or to attribute emissions to specific facilities in densely-equipped oil and gas basins~\cite{lauvaux2021global}.

High spatial-resolution hyper-spectral satellite imagery from PRISMA~\cite{cusworth2019potential} and GHGSat-C~\cite{first2021mckeever} offers much lower emission detection thresholds and the capacity to attribute precisely an emission to a specific oil and gas facility. Emissions as small as 0.2 t$_{CH_4}$/hr and 0.1 t$_{CH_4}$/hr have been detected by PRISMA and GHGSat-C instruments, respectively. However, the tasking nature and relatively small fields of view of these products limit their viability for persistent monitoring at a global scale. Airborne campaigns have an even better spatial resolution and lower detection limits (about 0.01 t$_{CH_4}$/hr). For example, AVIRIS~\cite{scafutto2021evaluation} has a detection threshold of the order of 0.01 t$_{CH_4}$/h, Scientific aviation's in-situ measurement offer a limit of detection below 0.005 t$_{CH_4}$/h \cite{amt-10-3345-2017}, the Kairos's passive imaging system~\cite{10.1525/elementa.2021.00063} has a wind speed-normalized detection limit of about 0.01 t$_{CH_4}$/hr for a wind speed of 1 m/s and the Bridger Photonics’ active system~\cite{JOHNSON2021112418} has an
absolute detection threshold on the order of 0.002 t$_{CH_4}$/hr at wind speeds of 3 m/s. However, airborne campaigns suffer from the same limited spatial coverage as the high spatial-resolution hyper-spectral satellite. 
The \emph{Sentinel-2} mission provides persistent multi-spectral imagery in the SWIR range and a two-to-ten-days revisit time. Although these instruments are not designed with methane detection in mind, it turns out that some of the bands are sensitive to its presence, thus enabling detection and quantification of large $\text{CH}_4$ emissions. It was shown by Varon~\etal~\cite{varon2021high} that combining the two SWIR bands impacted by methane increases the contrast of the plumes, and that having access to a reference image (at another date) without a $\text{CH}_4$ anomaly still improves this contrast. 

In this work, we first present an automatic quantification process for \emph{Sentinel-2}. This method is then validated using a specific event in the US. We show that detections from multiple different satellites (namely \emph{Sentinel-2}, \emph{Landsat-8} and \emph{Sentinel-5P}) are coherent and can be combined to improve the revisit time and provide a better monitoring. The existence of the event is validated thanks to airborne measurements from the Environmental Defense Fund (EDF).
We then applied our detection framework for large scale detection, quantification and uncertainty estimation of methane plumes using imagery coming from existing SWIR instruments onboard the \emph{Sentinel-2} and \emph{Landsat-8} satellites. The methodology was used to monitor oil and gas infrastructure in mainly three countries. This led us to detect about 1200 events, a dataset that we are making publicly available.

We combined our measurements with data from other instruments, \emph{Sentinel-5P} data presented by Lauvaux~\etal\cite{lauvaux2021global} more adapted to detect ultra-emissions as well as smaller events detected with airborne campaigns, one in California presented by Duren~\etal~\cite{duren2019california} and one in the Permian presented by Cusworth~\etal~\cite{cusworth2021intermittency}. Using this combination of observed emissions, we were able to validate the hypothesis proposed by Lauvaux~\etal\cite{lauvaux2021global} that a robust emission power law model exists. This shows that global observations of ultra-emitters ($>25$ t$_{\text{CH}_4}$/hr) serves as an indicator for the magnitude of many more unobserved events, at least in the range covered by \emph{Sentinel-} ($>2$ t$_{\text{CH}_4}$/hr) but potentially even lower. We also clustered the emissions based on their location, \ie by country, and emission frequency to have a better understanding of local behaviors and derive trends. This allows us to hint that the power law initially proposed by Lauvaux~\etal\cite{lauvaux2021global} is likely also valid starting from $0.1$ t$_{\text{CH}_4}$/hr.
The power law relationship has important implications for future monitoring systems. Our results suggest that one could only monitor the largest emitters and draw conclusions on smaller emitters, assuming that the slope is defined by structural variables (for example the ratio between small and large pipes) and by the maintenance operation procedures.
Moreover, it can be used to track progress by defining a reference year and comparing that year to the following ones.

\section{Principles for methane detection and quantification with multi-spectral satellite imagery}
\label{sec:principles}

\begin{figure}[t]
    \centering
    \includegraphics[width=.8\linewidth]{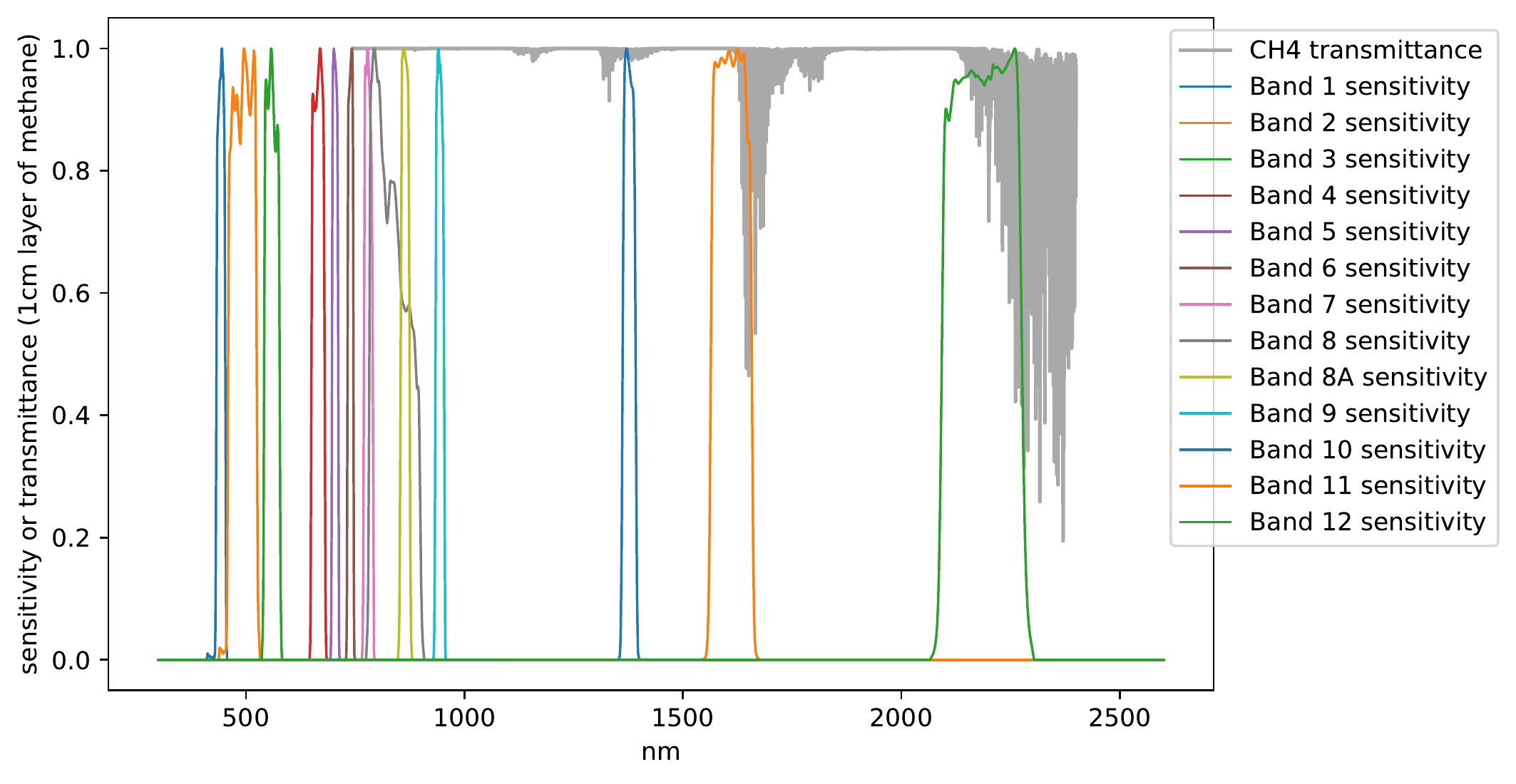}
    \caption{Methane transmittance spectrum and \emph{Sentinel-2 A} spectral sensitivity for all its bands. Ignoring the cirrus band B10 that is not suitable for monitoring applications, only bands B11 (1568-1659 nm) and B12 (2114-2289 nm) are impacted by the presence of methane in the atmosphere. The other bands are not impacted.}
    \label{fig:spectra}
\end{figure}

\begin{figure}[t]
    \centering
    \includegraphics[width=.8\linewidth]{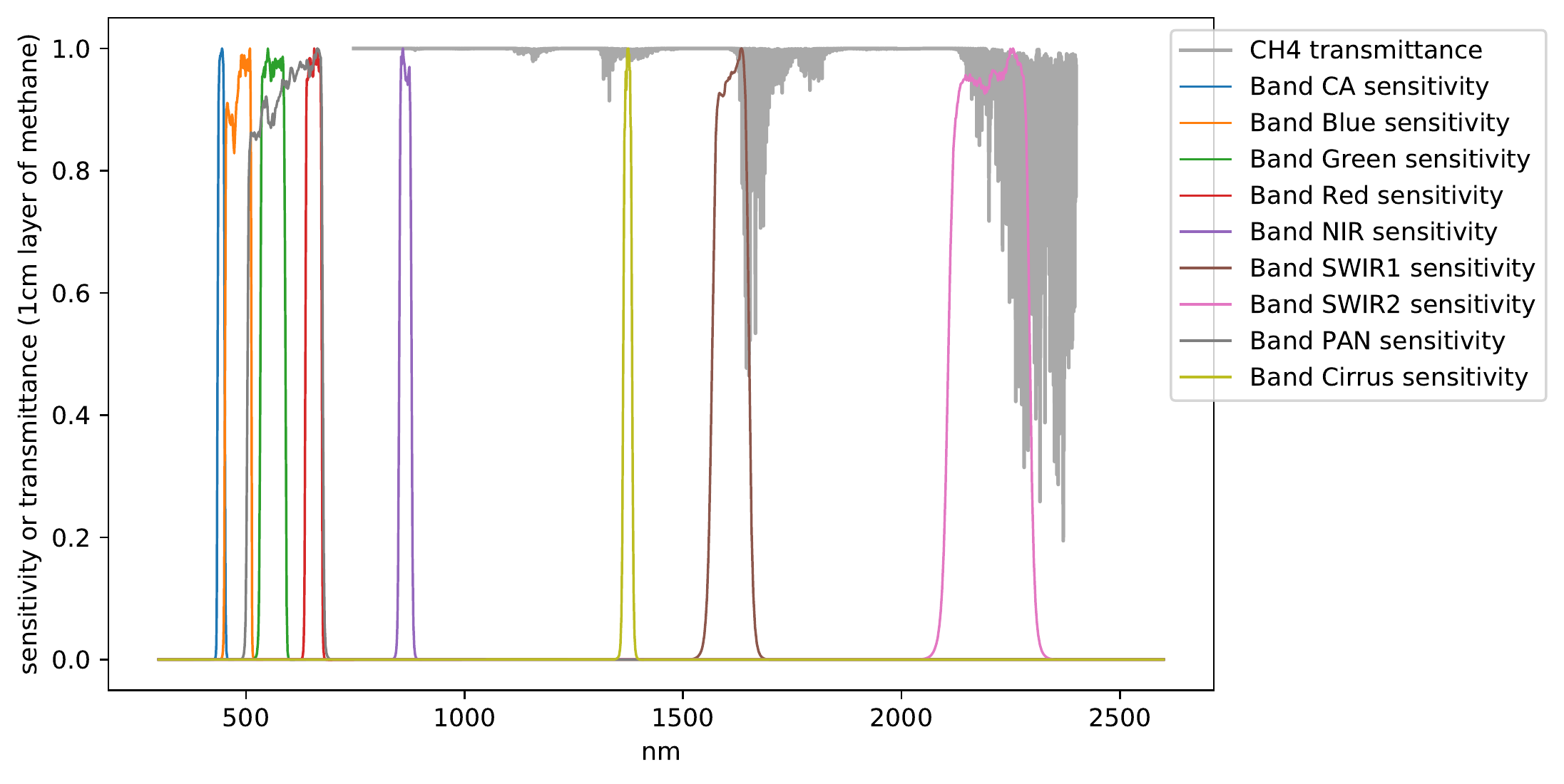}
    \caption{Methane transmittance spectrum and \emph{Landsat-8} spectral sensitivity for all its bands. Similarly to \emph{Sentinel-2}, the two SWIR bands SWIR1 and SWIR2 can be used for methane detection and quantification.}
    \label{fig:spectra_l8}
\end{figure}

When light traverses a gas, its intensity can be attenuated on certain wavelengths. Using this property, it is possible to detect the presence of a specific gas when its attenuation properties are known, and to derive a  quantification of the concentration of this gas. We apply this concept to methane detection using multi-spectral satellite imagery.
We focus on the detection and quantification of isolated excess concentrations of methane in the atmosphere, also referred as anomalies. These phenomena are often due to emissions in oil and gas infrastructures. 
Since methane absorbs light in the SWIR part of the spectrum, it is possible to use satellites such as \emph{Sentinel-2} (see Fig.~\ref{fig:spectra}) or \emph{Landsat-8} (see Fig.~\ref{fig:spectra_l8}) that provide a good spatial resolution, a low revisit time and free of charge.

\begin{figure}[t]
    \centering
    \includegraphics[width=.6\linewidth]{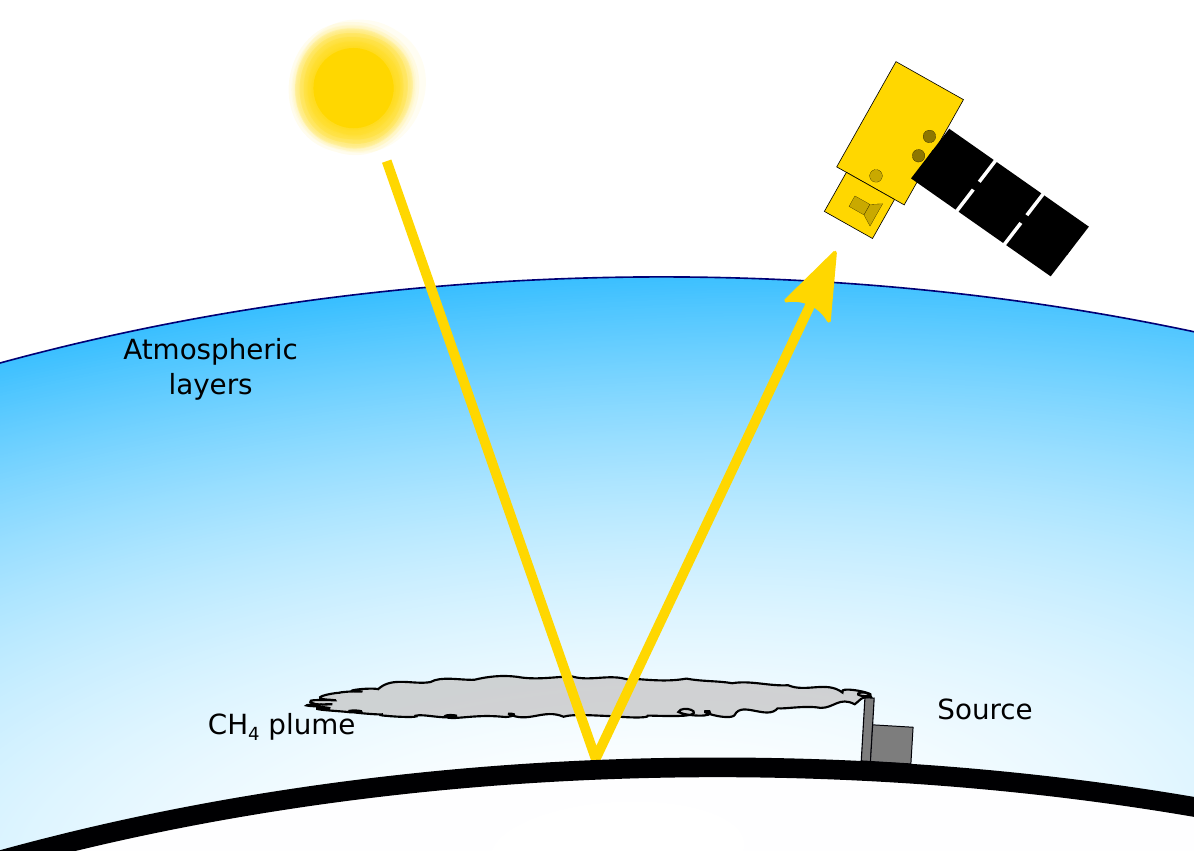}
    \caption{Atmosphere observation model. The on board sensor sees the light coming from the sun after reflection on Earth. Its intensity has dimmed due to the atmosphere and the reflection process. The dimming is also impacted by the presence of a methane plume.}
    \label{fig:acquisition}
\end{figure}

We use a simple absorption model to characterize the attenuation due to the presence of methane. The Beer-Lambert law states that for a light source with intensity $I_0$ and a wavelength $\lambda$
\begin{equation}
    I = I_0 e^{-\sum_{i=0}^N A_i(\lambda)l_i},
\end{equation}
where the light goes through $N$ gases defined by their absorption $A_i(\lambda)$ and equivalent optical path length $l_i$ defined as the product of the actual optical path and the concentration of the i$^{\text{th}}$ gas.
In our case, the $N$ gases correspond to the atmosphere and $I_0$ is the sunlight in the SWIR spectrum. We can also reasonably assume $I_0$ to be constant for all wavelengths $\lambda$ in each band respectively.
Taking into account that the sensor of a satellite integrates over a band of wavelengths described by a sensitivity function $s$, the intensity of the light seen by a space-borne sensor becomes 
\begin{equation}
    I = I_0 \int s(\lambda) \alpha(\lambda) e^{-\gamma \sum_{i=0}^N A_i(\lambda)l_i} d\lambda,
    \label{eq:atmospheric model}
\end{equation}
where the two passes through the atmosphere are taken into account in $\gamma$ (which is a function of both the sun azimuth angle and the satellite view angle).  The  reflection coefficient of the ground is represented in the formula by the surface albedo $\alpha(\lambda)$. See Fig. \ref{fig:acquisition}.

In the presence of a methane emission, characterized by $l_{leak}$, the intensity of the light seen by the sensors becomes
\begin{equation}
    I_{leak} = I_0 \int s(\lambda) \alpha(\lambda) e^{-\gamma \sum_{i=0}^N A_i(\lambda)l_i} e^{-\gamma A_{CH4}(\lambda)l_{leak}} d\lambda.
    \label{eq:atmospheric model_leak}
\end{equation}
Supposing that we have both the exact same observation with and without a methane emission, it becomes very easy to detect the emission. Indeed $I_{leak} < I$ everywhere $l_{leak}$ is non zero. 
The problem is that the observation without methane, also called background observation, is never available in practice. Therefore, a reference observation without methane is needed in order to distinguish an attenuation due to the presence of methane from a difference in the surface albedo.

When we assume that methane emissions are anomalous events, it is to be expected that most observations in a time series should not contain excess methane. 
So, if we suppose that the surface albedo is rather stable in time, the time series can be used to estimate a methane free background model that can be compared with the current observation. 
Here, we compute the background for a given date as its linear regression over the previous dates. If we denote by $I_t$ the observation at time $t$, then the regression computes the optimal   weights $w_i$ that solve 
\begin{equation}
      \min_{ \{w_i\} } \left\|I_t - \sum_{i=0}^{t-1}  w_i I_i \right\|^2.
\end{equation}
Then the background is obtained as the linear combination $\sum_{i=0}^{t-1}  w_i I_i$.
To further improve the background subtraction we combine this estimation with a band ratio that exploits the correlation between SWIR bands, similarly to the \emph{multiple-band single-pass} (MBSP) from Varon~\etal~\cite{varon2021high}.

Quantifying emissions is also an important part of monitoring. While the previous processing was presented for methane emission detection, it can also be used to quantify it. Supposing that both the signal with the emission $I_{leak}$ and without emission $I_{bg}$ are available (using for example the process presented previously), then 
    \begin{equation}
       \frac{I_{leak}}{I_{bg}} \approx \frac{\int_{B12} s(\lambda) e^{-\gamma \sum_{i=0}^N A_i(\lambda)l_i} e^{-\gamma A_{CH4}(\lambda) l_{leak}} d\lambda}{\int_{B12} s(\lambda) e^{-\gamma \sum_{i=0}^N A_i(\lambda)l_i} d\lambda}. 
    \end{equation}
Since $\gamma$ is known for each acquisition, this ratio only depends on the atmosphere composition. Therefore, for a fixed atmosphere composition, it is possible to estimate the value of $l_{leak}$ as the solution of a simple optimization problem
    \begin{equation}
    \label{eq:quantif}
       \argmin_{l_{leak}} \left\|\frac{I_{leak}}{I_{bg}} - \frac{\int_{B12} s(\lambda) e^{-\gamma \sum_{i=0}^N A_i(\lambda)l_i} e^{-\gamma A_{CH4}(\lambda) l_{leak}} d\lambda}{\int_{B12} s(\lambda) e^{-\gamma \sum_{i=0}^N A_i(\lambda)l_i} d\lambda} \right\|^2_2. 
    \end{equation}
In practice, the atmosphere model can be well approximated with a simple ``pure methane atmosphere", \ie an atmosphere that's purely made of methane, instead of considering a complete atmosphere model. 
This quantification scheme can also be adapted when using band ratio.

\section{Practical methane emission tracking}
\label{sec:pratical}

\begin{figure}[t]
    \centering
    \begin{tabular}{lll}
    \includegraphics[height=.26\linewidth]{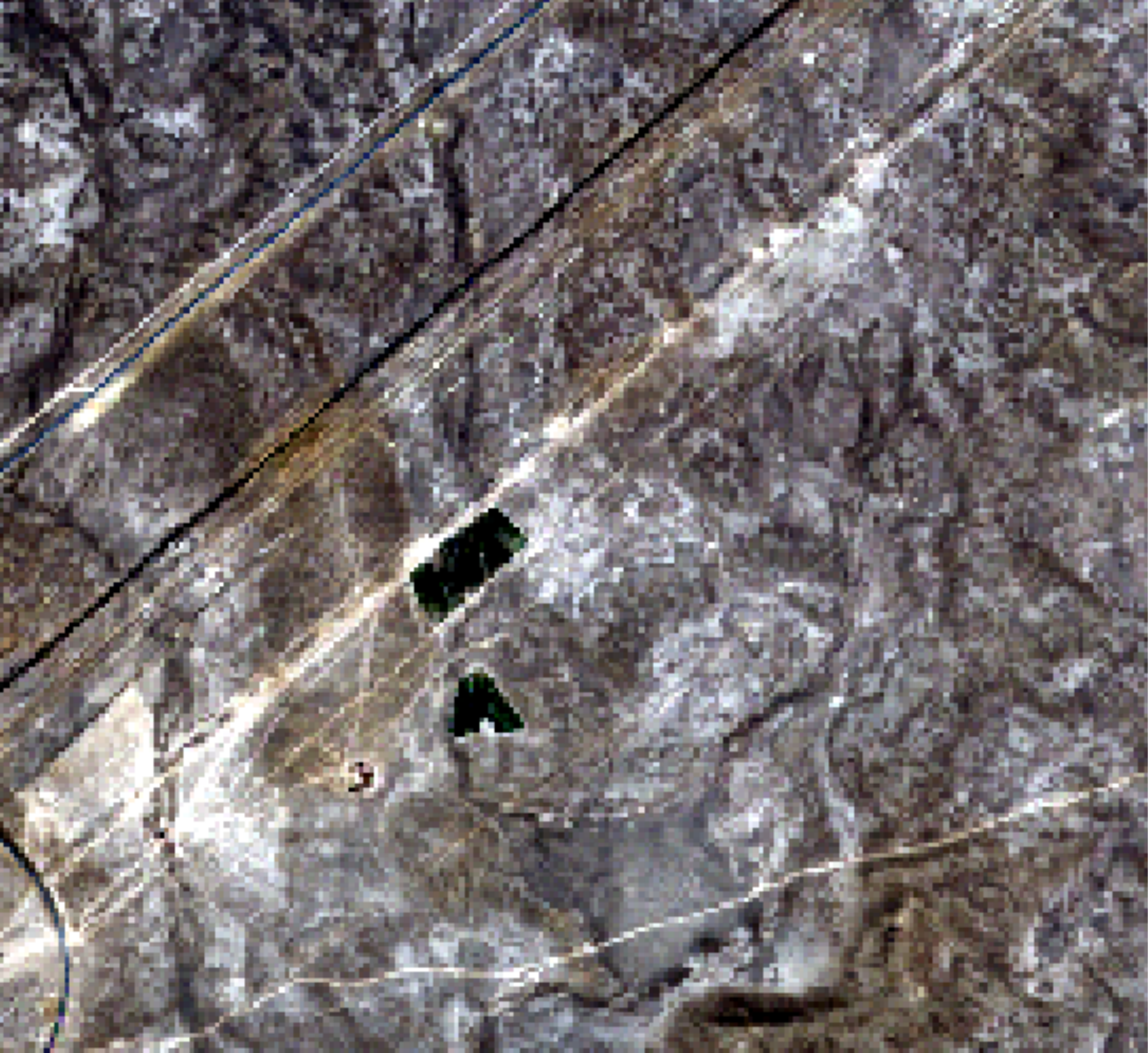}&
    \includegraphics[height=.26\linewidth]{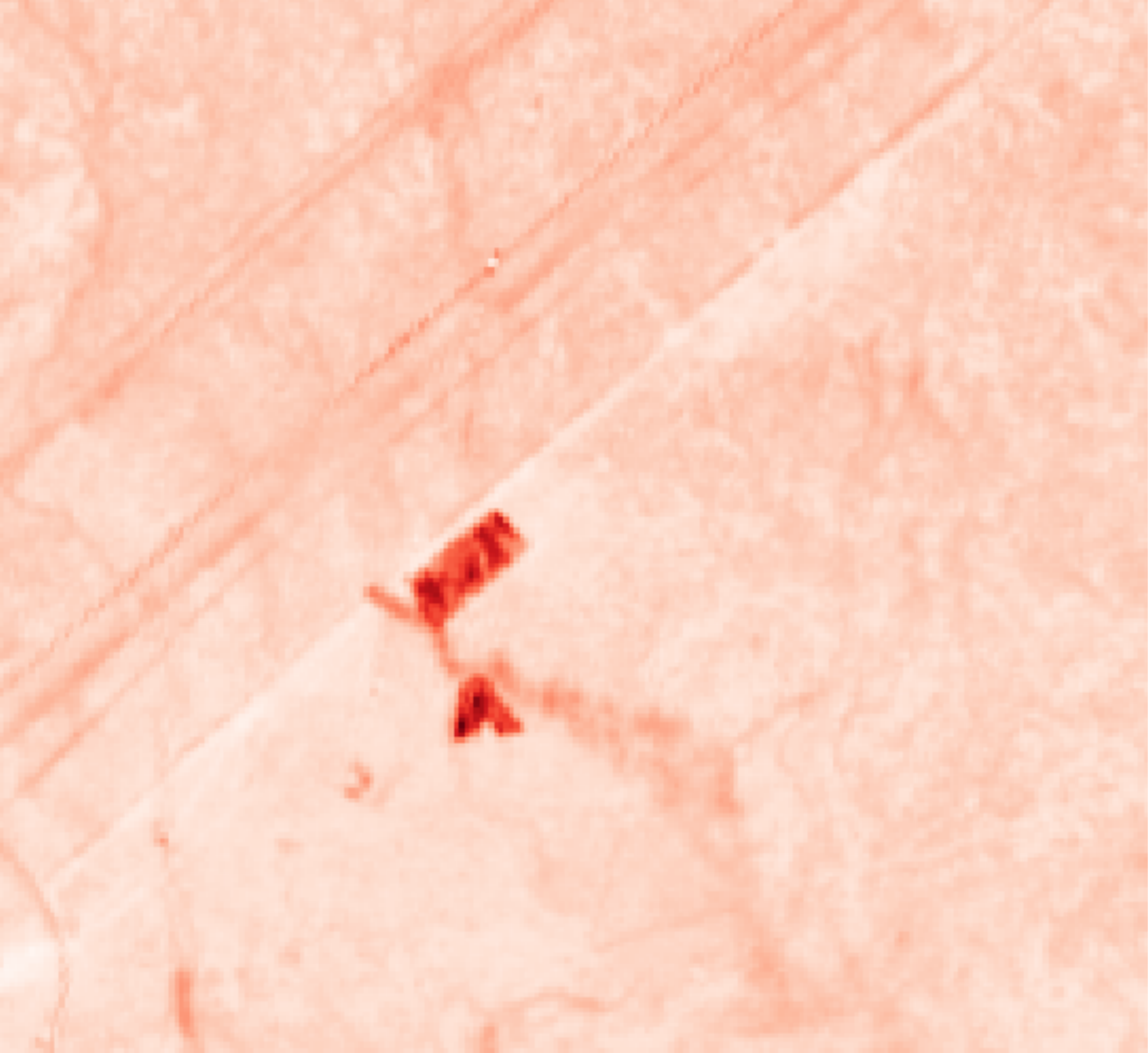}&
    \includegraphics[height=.26\linewidth]{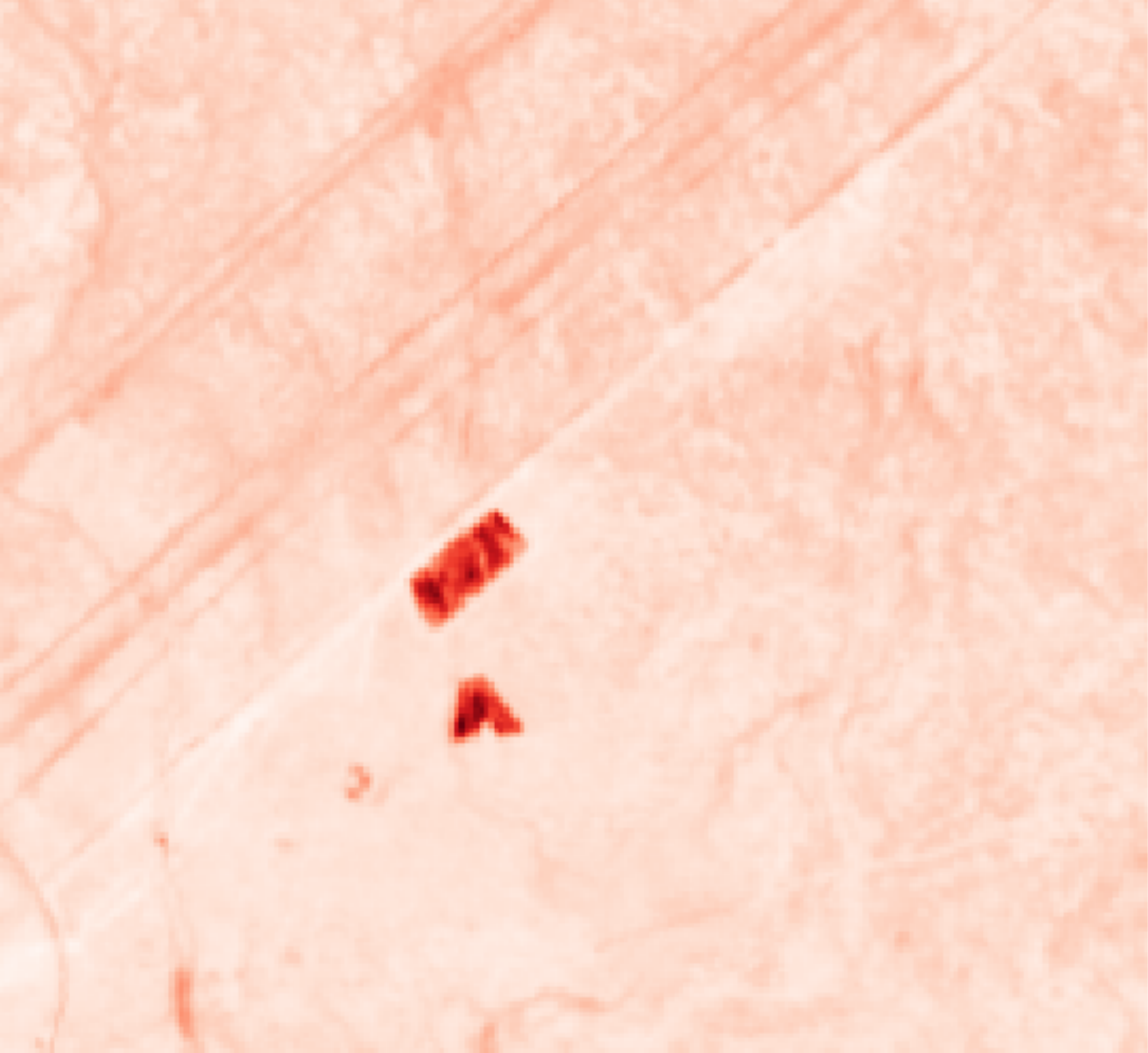}\\
    \includegraphics[height=.26\linewidth]{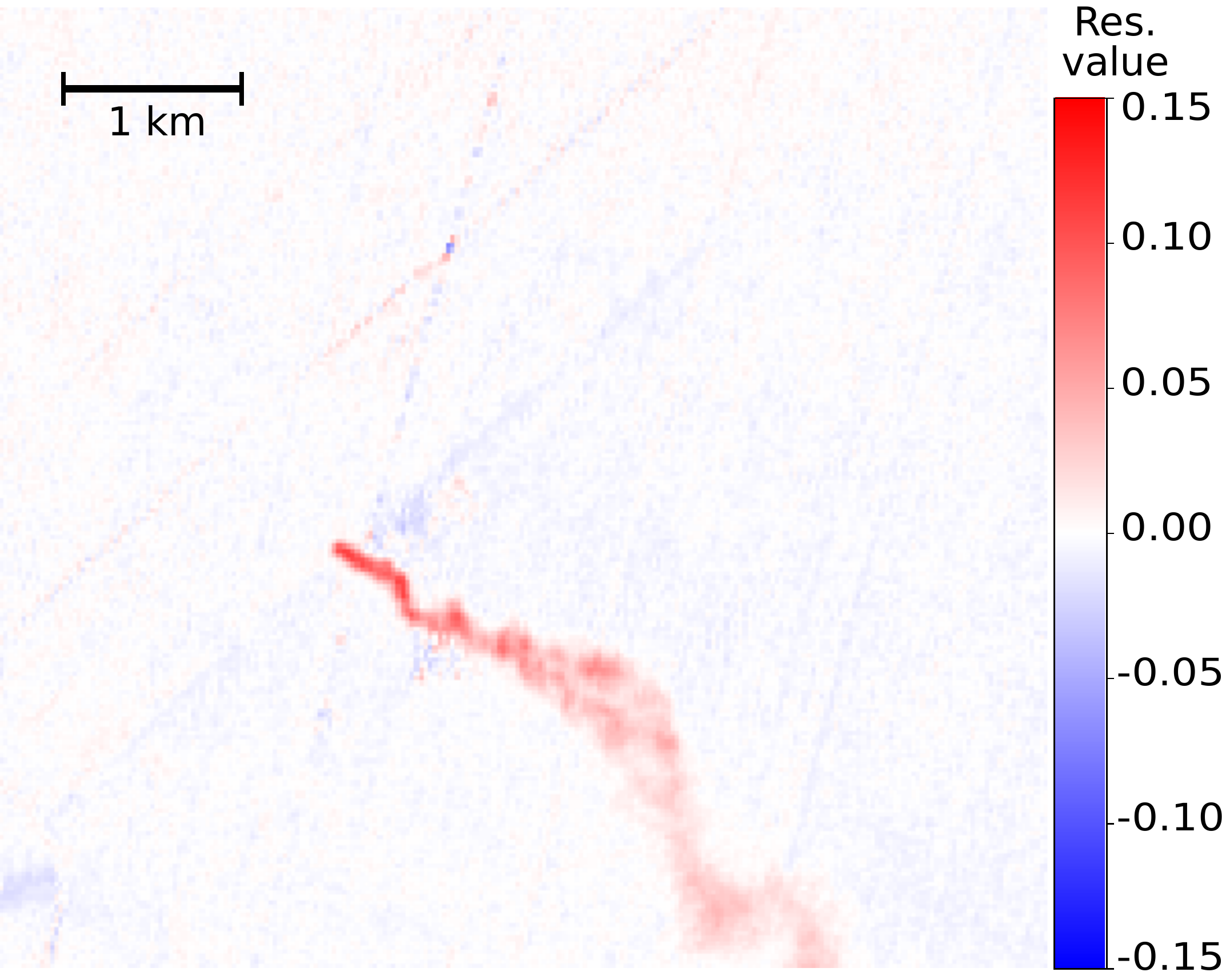}&
    \includegraphics[height=.26\linewidth]{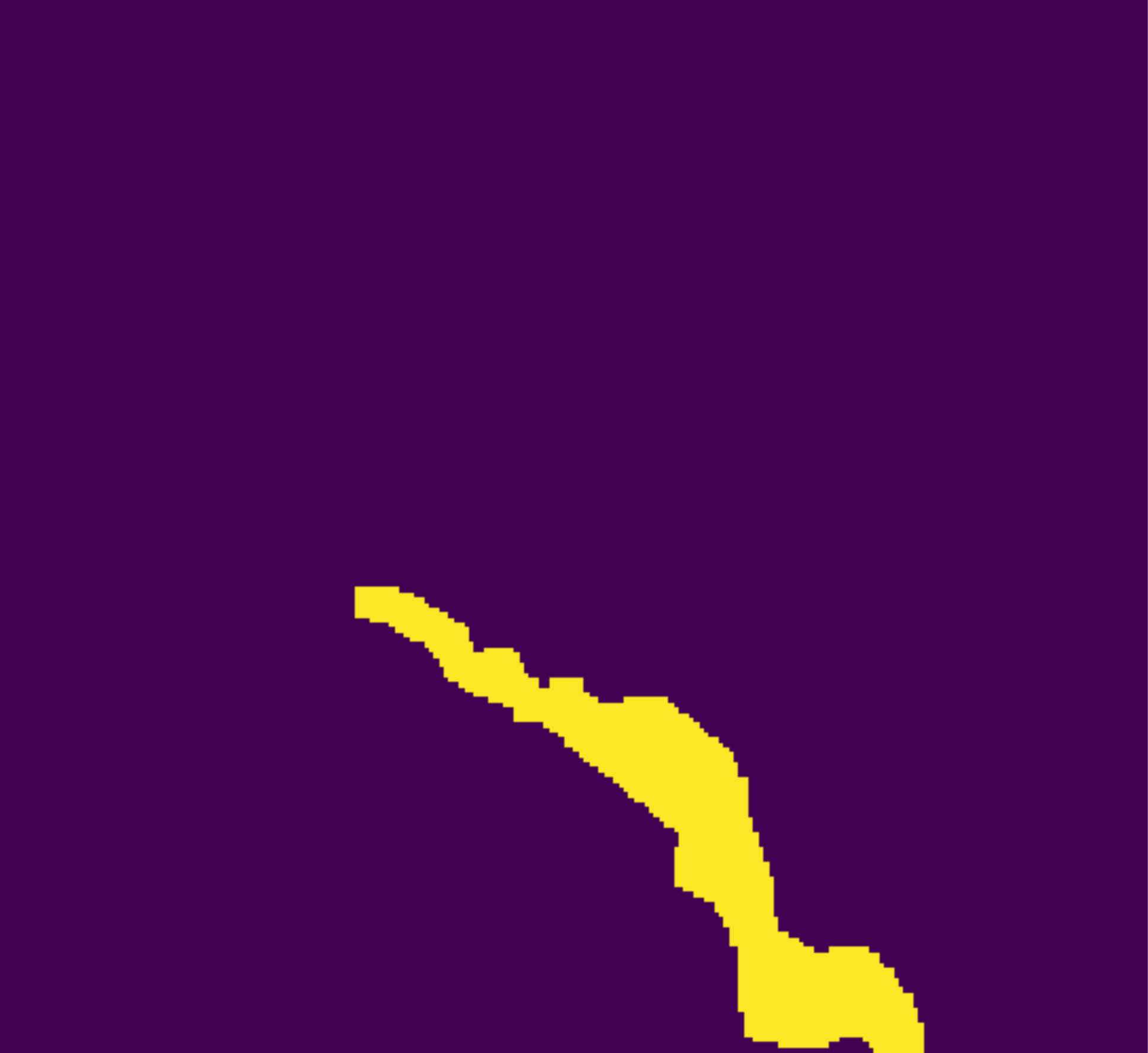} &
    \includegraphics[height=.26\linewidth]{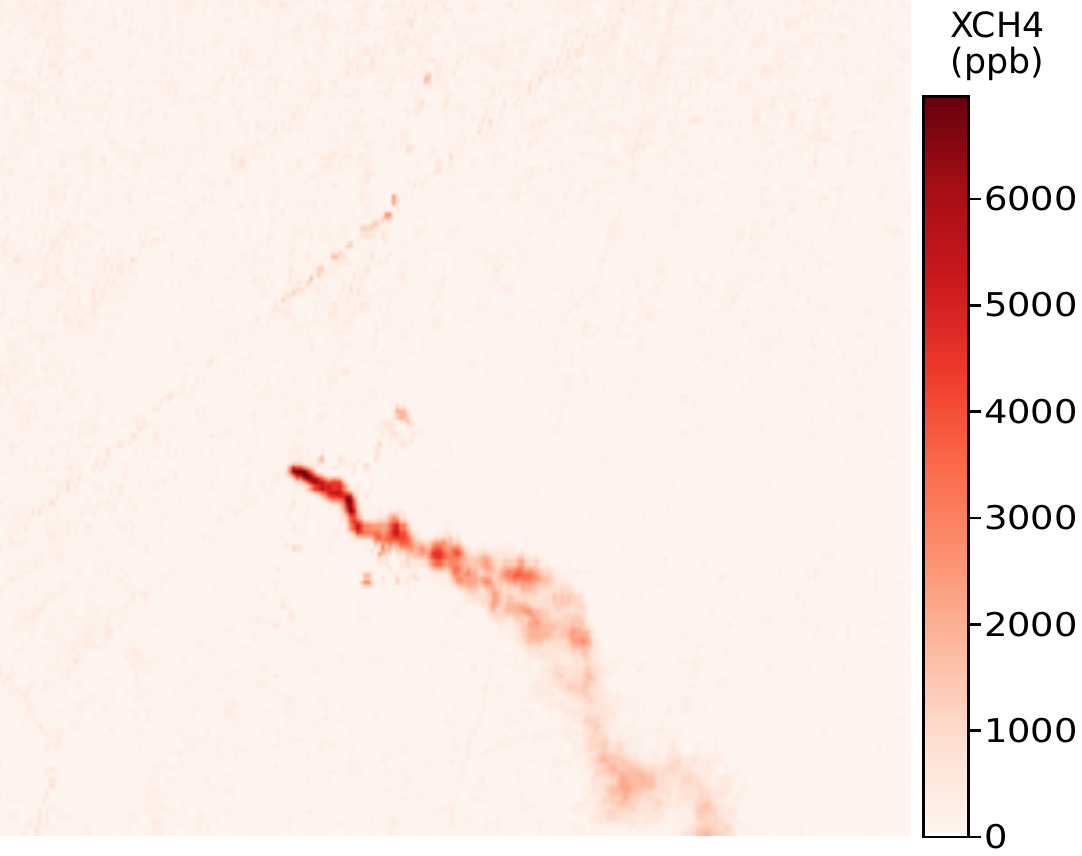} 
    \end{tabular}
    \caption{Visualization of each step of the detection and quantification method. From left to right, top to bottom: the \emph{Sentinel-2} image of the location (only the RGB channels are shown), the log band ratio corresponding to the image, the predicted background model, the residual showing the plume, the mask corresponding to the detected plume and the associated quantification in ppb.}
    \label{fig:example_plume}
\end{figure}

\begin{figure}[t]
    \centering
    \begin{tabular}{lll}
     \includegraphics[height=.26\linewidth]{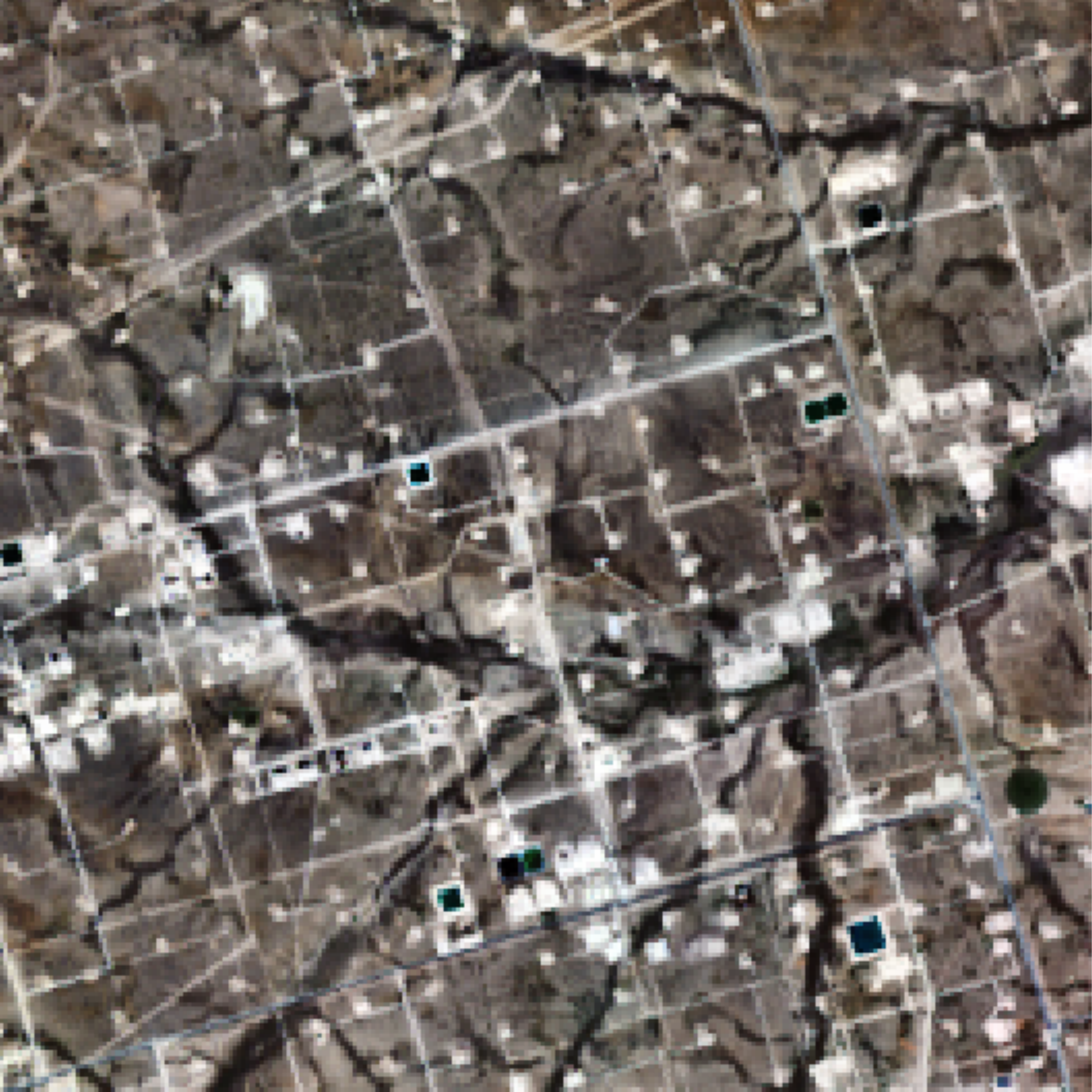}&
    \includegraphics[height=.26\linewidth]{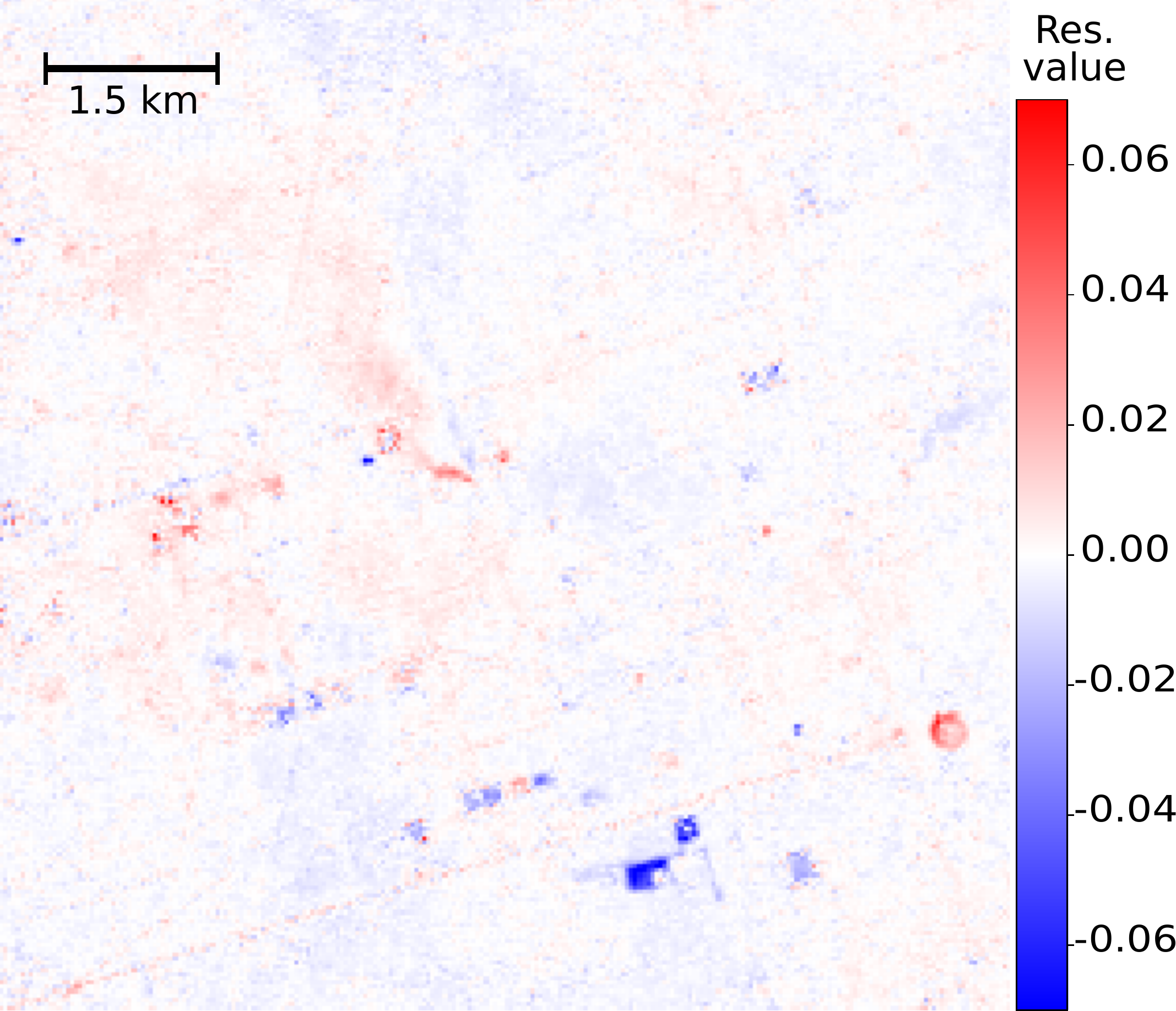}&
    \\
    
    \includegraphics[height=.26\linewidth]{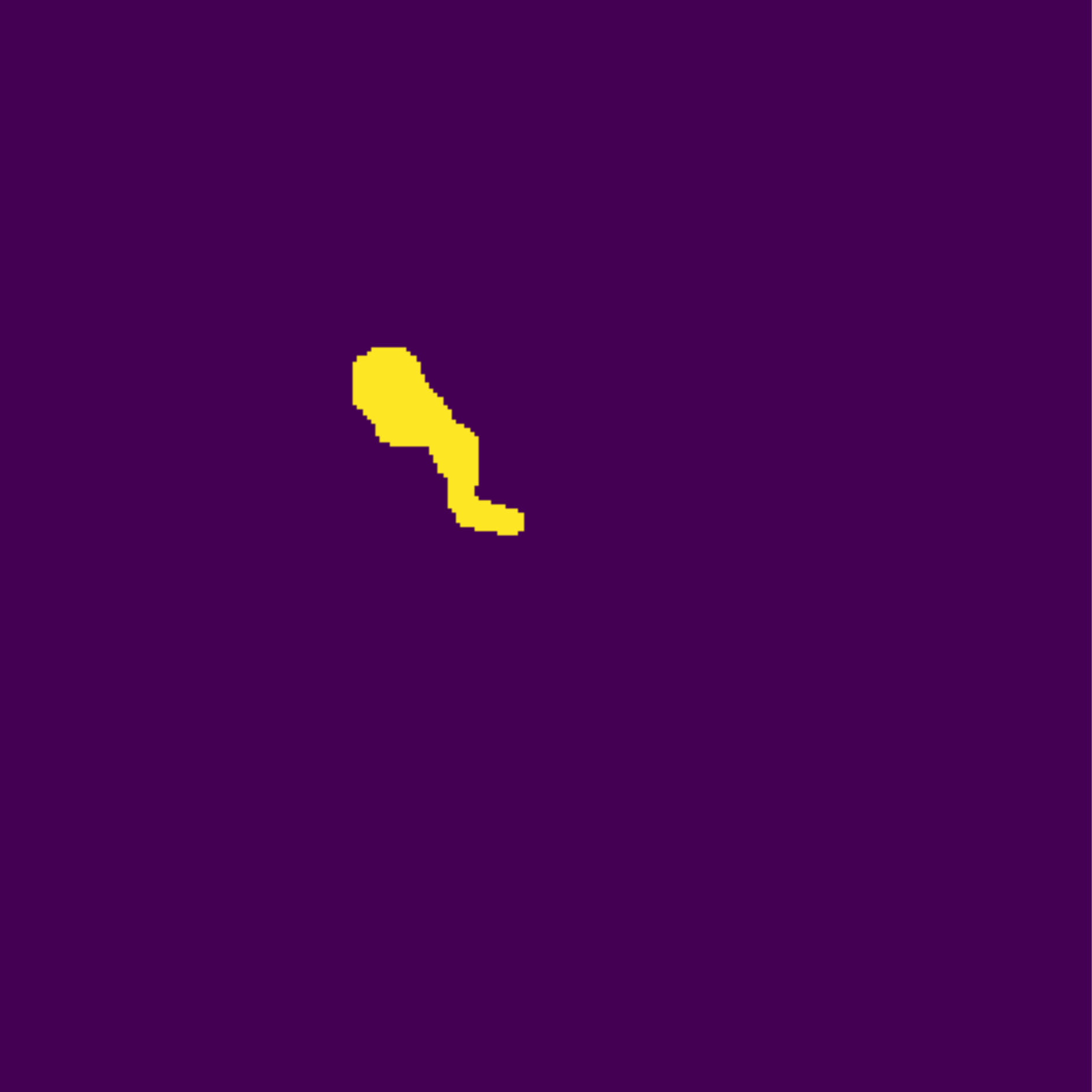} &
    \includegraphics[height=.26\linewidth]{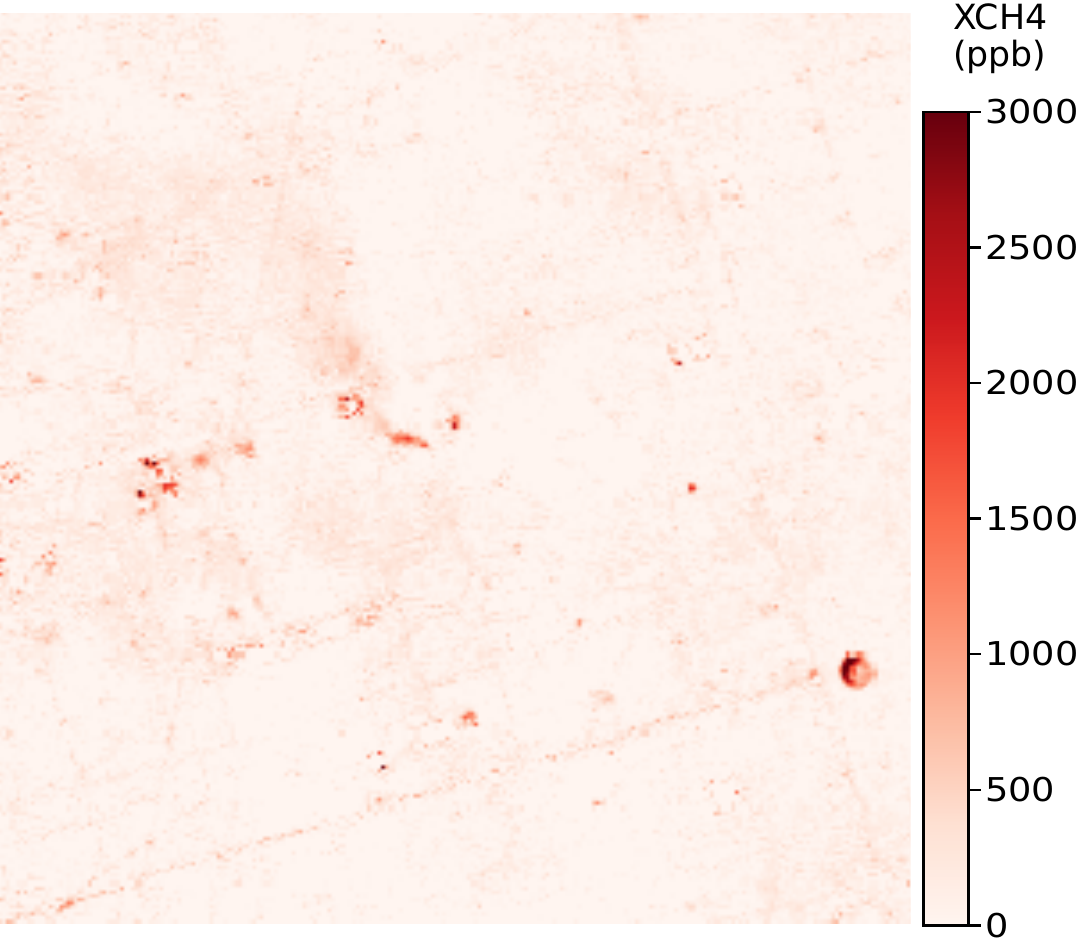} 
    \end{tabular}
    \caption{Visualization of each step of the detection and quantification method. From left to right, top to bottom: the \emph{Landsat-8} image of the location (only the RGB channels are shown), the log band ratio corresponding to the image, the residual showing the plume, the mask corresponding to the detected plume and the associated quantification in ppb.}
    \label{fig:example_plume_l8}
\end{figure}

We present in this Section the practical implementation of the detection and quantification principles mentioned in the previous section. Namely, we first present the different preprocessing steps necessary for the method to function. We provide more details about the background reconstruction process, the detection validation process and the quantification process. This practical methodology is the one used to perform all the experiments presented in this paper. Fig.~\ref{fig:example_plume} illustrates the different steps of the proposed methodology for \emph{Sentinel-2}; Fig.~\ref{fig:example_plume_l8} illustrates the same steps but for \emph{Landsat-8}. 


\subsection{Preprocessing}
\label{sec:practical:preprocessing}

\begin{figure}[t]
    \centering
    \begin{subfigure}{.32\linewidth}
    \includegraphics[trim={0 0 300 300}, clip, width=\linewidth]{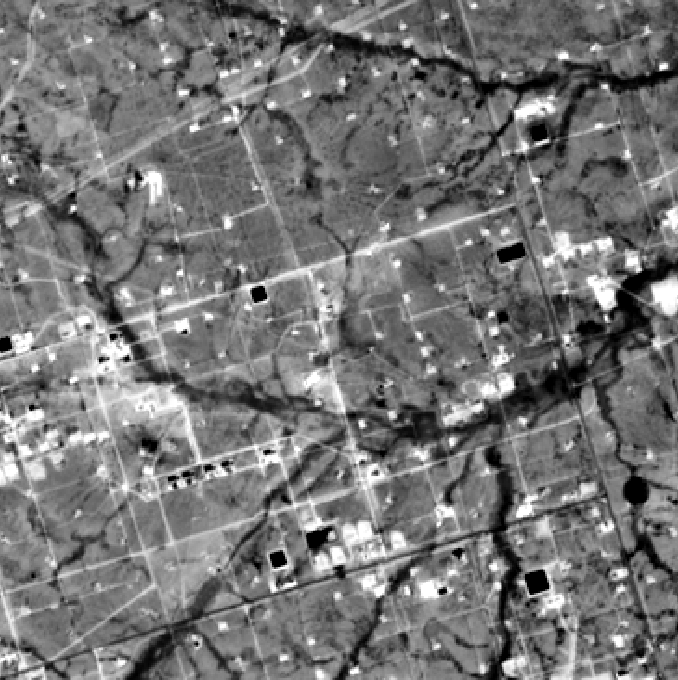}
    \caption{Band 12 of \emph{Sentinel-2} \phantom{to align everything}}
    \end{subfigure}
    \begin{subfigure}{.32\linewidth}
    \includegraphics[trim={0 0 300 300}, clip, width=\linewidth]{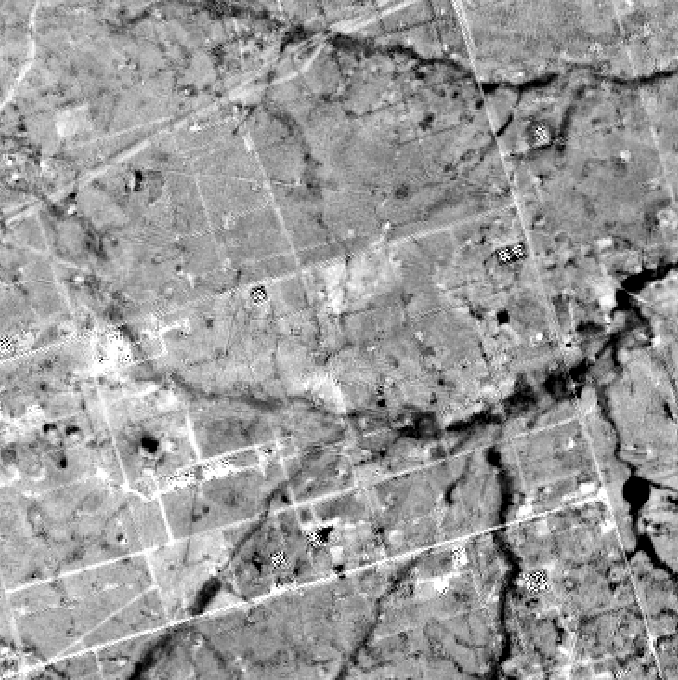}
    \caption{Ratio between B12 and 11 without an antialiasing filter}
    \end{subfigure}
    \begin{subfigure}{.32\linewidth}
    \includegraphics[trim={0 0 300 300}, clip, width=\linewidth]{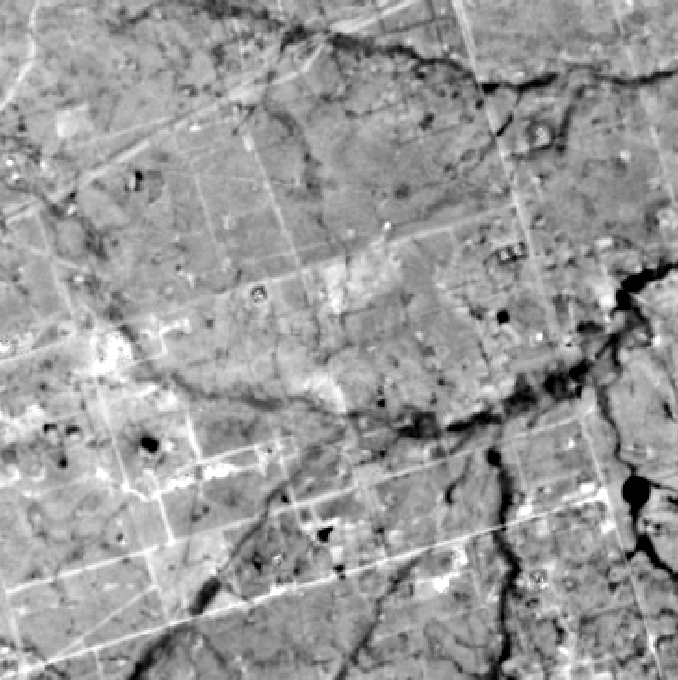}
    \caption{Ratio between B12 and 11 with an antialiasing filter}
    \end{subfigure}
    \caption{Eventhough it is barely visible, \emph{Sentinel-2} images are aliased and therefore should be preprocessed before computing the ratio. The artifacts due to aliasing would impact the processing otherwise. A Gaussian filter with parameter $\sigma=0.7$ allows to remove this aliasing before the processing. Figure best seen zoomed.}
    \label{fig:example_alias}
\end{figure}

From now on, we consider areas of interest of size approximately 10x10 km$^2$. We found out that this size is well adapted to capture methane plumes created by emissions, while being large enough so that the reconstruction is not impacted too much by the presence of methane in the reference images. We collect L1C \emph{Sentinel-2} timeseries corresponding to the areas of interest, preferably considering timeseries longer than six months. We first co-register all the images of a timeserie using the method by Hessel~\etal~\cite{hessel2021global}. We also apply a cloud detection algorithm, such as the one proposed by Dagobert~\etal~\cite{ipol.2020.271}, to estimate the cloud cover. All images with more than  $15\%$ of the pixels covered by clouds are discarded.
Sentinel-2 images comprise $12$ bands with spatial resolutions from $10$m per pixel to $60$m per pixel. The two bands of interest, namely band 11 and band 12, are both sampled at $20$m per pixel therefore there is no need for resampling them. However, we have observed that these two bands are aliased. This is particularly important because we are computing ratios of these two bands and therefore this aliasing can create large artifacts during the processing (see Fig.~\ref{fig:example_alias}). In order to avoid this problem we apply an anti-aliasing filter, namely a Gaussian filter with parameter $\sigma=0.7$, prior to any other processing. We also apply a log on the ratio. This limits the impact on the reconstruction of abnormal high values present in the SWIR bands, for example due to flaring, 
{which are frequently found in the vicinity of oil and gas facilities}.
As it will be seen in the \emph{Detection validation} subsection, the other bands are still useful to validate a plume detection. This is why we resample all these bands to $20$m per pixel so that comparison is easier.

\subsection{Background estimation}
\label{sec:practical:reconstruction}

The core of the detection method is the background estimation process. For that, each time-series of log band ratios computed above is processed using a sliding window of size $T=30$ dates. For each date in the window, we compute its linear projection on the past $T-1$ images. Using the estimated background, we define a residual that corresponds to the difference between the input data and the prediction. A longer time series improves the SNR of the extracted plume, thus fostering its detection~\cite{machover_eureka_patent2020}.
Note that by projecting on a time series there is no need to manually choose a reference date as background~\cite{varon2021high}.

\begin{figure}[t]
    \centering
    \includegraphics[width=.99\linewidth]{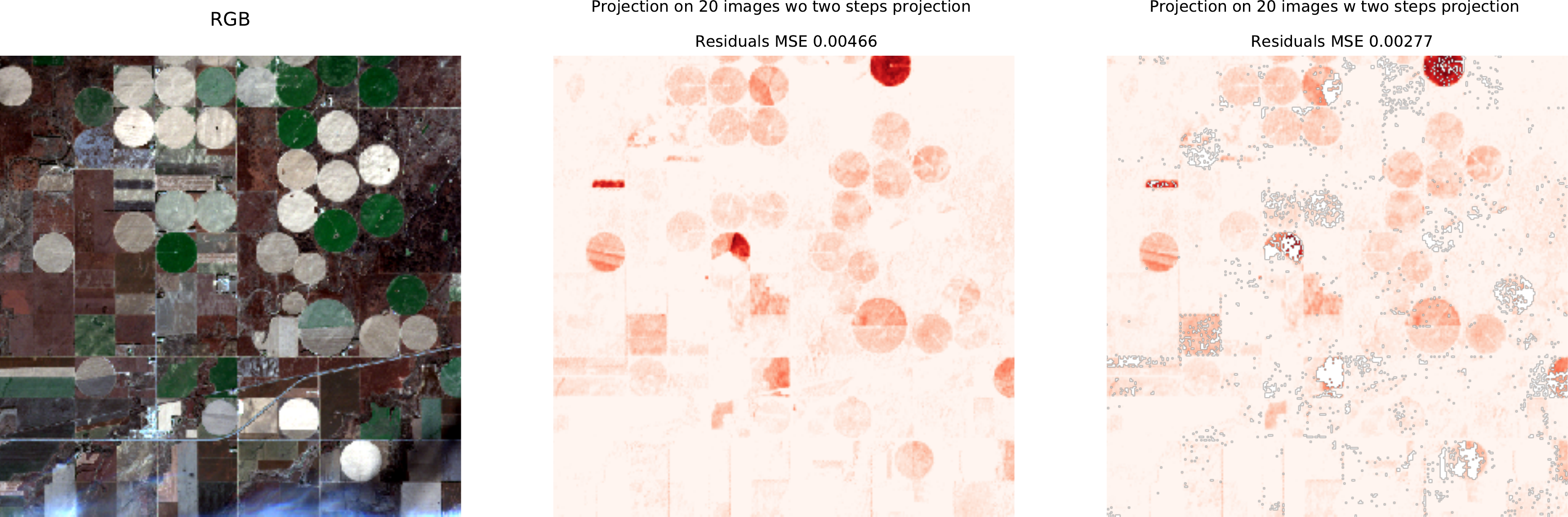}
    \caption{Improvement of background estimation when using a two-steps estimation. Left to right: The RGB image of the scene, the background estimated using a single step, the background estimated using two steps. Using two steps allows to have a much smaller MSE than a single step estimation (with the outliers set to zero). The MSE has been estimated on the same pixels (\ie without the outliers) for both images.}
    \label{fig:two stage}
\end{figure}

Similarly to how flaring could impact the background estimation, new (or disappearing) large structure can also lead to errors in the quantification. To limit the impact of outliers, robust estimation methods such as Huber regression~\cite{huber2004robust} or the iteratively reweighted least square algorithm~\cite{weiszfeld1937point} can be used. We found out that in our case   such robust regression methods are quite slow. For this reason, we use an approximate two-steps estimation method that is good enough for this application.
A first estimation is done using a linear projection as presented previously. Then the $5\%$ of pixels with the worst estimation are discarded. The remaining pixels are then used to perform a second linear projection, this time without the outliers. The coefficients estimated with the second linear projection are used to perform the final estimation. We argue that even if pixels containing methane are initially discarded, this is not a problem because methane should, by definition, not impact the background prediction. Fig.~\ref{fig:two stage} shows a case in which this procedure allows to refine the background reconstruction. The reconstruction error of the background  is almost twice as small when using a two step estimation.

\begin{figure}[t]
    \centering
    \includegraphics[width=.99\linewidth]{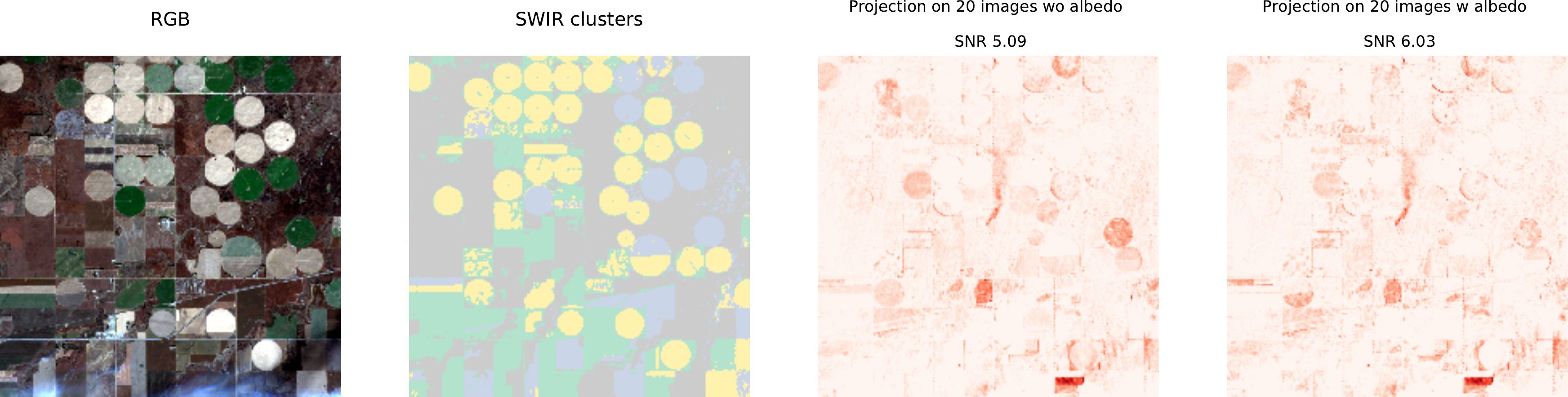}
    \caption{Improvement of background estimation when using clustering. Left to right: The RGB image of the scene, the different clusters estimated from the albedo, the background estimated without the clusters, the background estimated with the clusters. Using a clustering step during the background estimation increases the SNR of the methane plume with respect to the background.}
    \label{fig:subtraction with clusters}
\end{figure}

Despite removing outliers, the two-stage approach cannot deal with time series containing large zones with changing albedo. This is the case for the crop fields  seen in Fig.~\ref{fig:subtraction with clusters}, for which a spatially adaptive processing must be adopted.
The objective is to bring non linearity to the projection by performing one projection for each zone of similar albedo. An albedo map is computed by clustering the pixels of our images with four different features: the temporal standard deviation and median of the absorbing band, the $x$ and $y$ position of the pixel in the image. The clustering is done using a Gaussian mixture model, and the optimal number of clusters is fixed with the post analysis of the Bayesian information criterion of the clustering~\cite{schwarz1978estimating}. This methodology being more computationally intensive is performed only on regions with a high albedo variance such as regions with many crop fields as shown Figure~\ref{fig:subtraction with clusters}.

\subsection{Detection validation}
\label{sec:practical:validation}

While we would like to have a completely automatic detection process, directly detecting on the residuals computed in \emph{Background estimation} yielded too many false detections. This is why we added an extra step where all detections are done and verified manually. 
In particular, a mask corresponding to the shape of the potential plume is first manually annotated. We then compare the content of the annotated region to the content of the same region but in the other bands. If the potential plume is indeed a true methane plume, then it should not be correlated to the content of the bands that are not impacted by the presence of methane. In particular, a similar shape should not be found in these other bands. 
Some surfaces, for example snow, have a higher reflectance in B11 than B12. This causes a contrast inversion and a dimming-like phenomenon when looking at the band ratio.  
Because of that, it is possible that potential plumes appear in the band ratio even though they do not correspond to an actual dimming in B12. 
The last validation step checks that the detection corresponds indeed to a dimming in B12.

\subsection{Source quantification}
\label{sec:practical:quantification}

Once the mask of the plume is available, we quantify the emission rate corresponding to the source of the plume. The first step is to quantify the equivalent amount of methane $l_{CH4}$ per pixel that corresponds to the extra methane attributed to the source. For that, we adapt the quantification model presented in Eq.~\ref{eq:quantif} so as to take into account the extra log preprocessing as well as the band ratio. This leads to an excess of methane $\Delta\Omega(p)$ for the pixel $p$ corresponding to
    \begin{equation}
    \footnotesize
        \Delta\Omega(p) = \argmin_{l_{leak}} \left\|R(p) - \log\left(\frac{\int_{B12} s(\lambda) e^{-\gamma A_{CH4}(\lambda) (l_{atm} + l_{leak})} d\lambda}{\int_{B12} s(\lambda) e^{-\gamma A_{CH4}(\lambda) l_{atm})} d\lambda} \frac{\int_{B11} s(\lambda) e^{-\gamma A_{CH4}(\lambda) l_{atm})} d\lambda}{\int_{B11} s(\lambda) e^{-\gamma A_{CH4}(\lambda) (l_{atm} + l_{leak})} d\lambda}\right)\right\|^2_2,
    \end{equation}
    with $R(p)$ the estimated residual at pixel $p$ and $l_{atm}$ the amount of methane naturally present in the atmosphere. We define $l_{atm}$ such that it corresponds to a residual background of $1800$ppb of methane. In practice, while the specific background level might fluctuate depending on the location and time, the error due to this approximation for the estimation of the excess of methane is negligible compared to all other sources of uncertainty (see~\cite{ehret2022automatic}). To estimate $A_{CH4}$, we use the HITRAN database~\cite{HITRAN2016}. We also use the sensitivity $s$ of \emph{Sentinel-2 A}, respectively \emph{Sentinel-2 B}, calibrated in laboratory provided by ESA %
(\url{https://sentinels.copernicus.eu/web/sentinel/technical-guides/sentinel-2-msi/performance}). 
The optimization is done using the downhill simplex algorithm.

Once each pixel of the plume has been quantified, we estimate the source emission rate using the integrated mass enhancement (IME) method~\cite{varon2018quantifying}. The IME method relates the source rate $Q$ to the total detected plume mass by
\begin{equation}
    Q = A \frac{U_{eff}}{L} \sum_{p \in \mathcal{M}}^{} \Delta \Omega(p),
\end{equation}
where $U_{eff}$ corresponds to the effective wind speed, $L$ the plume length, $\mathcal{M}$ the mask of the plume, $A$ the area covered by a pixel (in this case $A=400\text{m}^2$). Similarly to Varon~\etal~\cite{varon2018quantifying}, we estimated $L$ using the plume mask $M$ such that $L = \sqrt{A \times \#M}$ with $\#M$ the number of pixels in the plume mask $M$.
We use wind data collected from the ECMWF-ERA5 reanalysis product from the Copernicus Climate Change Service~\cite{ERA5}. Varon~\etal~\cite{varon2018quantifying} showed that $U_{eff}$ can be related to the local wind speed at 10m $U_{10}$ therefore we select the wind product at 10m above ground level and at the closest time before the sensing time for each estimation.
The source origin is selected manually using jointly the wind data and the plume shape.

\begin{figure}[t]
    \centering
    \includegraphics[width=.902\linewidth]{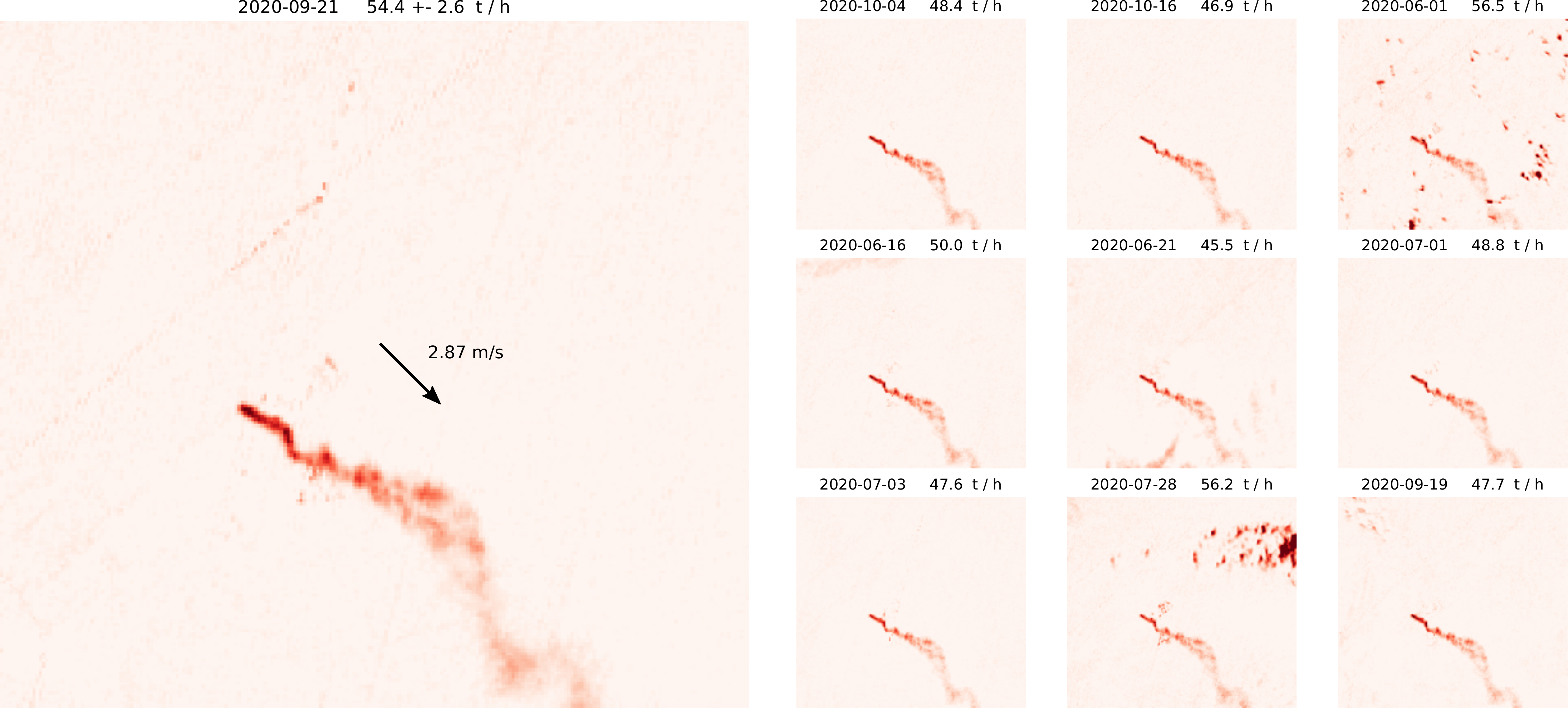}
    \includegraphics[trim={264 0 0 0},clip, width=.075\linewidth]{fig/fig_ppb.pdf}
    \caption{The uncertainty of the plume on the left is estimated by simulating the same plume in other images of the timeseries. Each estimated plume (in ppb) as well as their predicted emission rate are shown on the right. The arrow corresponds to the wind information used for the emission rate quantification.} 
    \label{fig:simulation}
\end{figure}

Note that, in the following, \emph{Sentinel-5P} measurements are not estimated using this process. They are instead derived from the methane concentrations provided by the Level-2 methane product. For that, we estimate a background methane concentration, by computing the median methane concentration neighboring a plume in the Level-2 product, which we remove from the measurements so that only the excess methane is measured.

\subsection{Quantifying the uncertainty caused by the proposed background removal method}

Different factors can contribute to quantification errors in the proposed method.
We focus here on the uncertainty induced by the proposed background 
estimation method by providing a per-scene uncertainty estimation. 
Note that the error estimated here does not include the uncertainty due to the IME process, the two need to be combined to obtain the final uncertainty corresponding to the source emission.

Fluctuations in the albedo and atmospheric conditions might be wrongly quantified as excess of
methane. The idea is to estimate the quantification errors due to these fluctuations by simulating the same methane plume
in other images of the time series: 
From each new simulated image the quantification method is run again and a new emission 
rate is estimated.
The uncertainty is then obtained as the standard deviation of the emission rates 
estimated with the simulated images. Fig.~\ref{fig:simulation} illustrates the concentrations obtained 
by applying this procedure on different images of a time series.

\section{Validation of the proposed methodology: the September 2020 Permian event}
\label{sec:permian_event}

We applied the proposed methodology to estimate emission rates during an event in the Permian basin. This event occurred during the summer of 2020 %
(estimated latitude and longitude of the source:  (31.7335, -102.0421))
and lasted about two months. Several observations from \emph{Sentinel-2}, \emph{Landsat-8}, \emph{Sentinel-5P} were collected and airborne hyperspectral observations from previous campaigns are available. 
The airborne measurements were obtained in September 2020 (towards the end of the event) with Scientific Aviation flights and are provided by the PermianMap project  (Operator Performance Dashboard: %
data from U. Arizona, NASA-JPL, and EDF provided via the PermianMap project by EDF (\url{https://data.permianmap.org/pages/operators}). Users are bound by the Terms of Use of this data).

\begin{figure}[t]
    \centering
    \includegraphics[width=\linewidth]{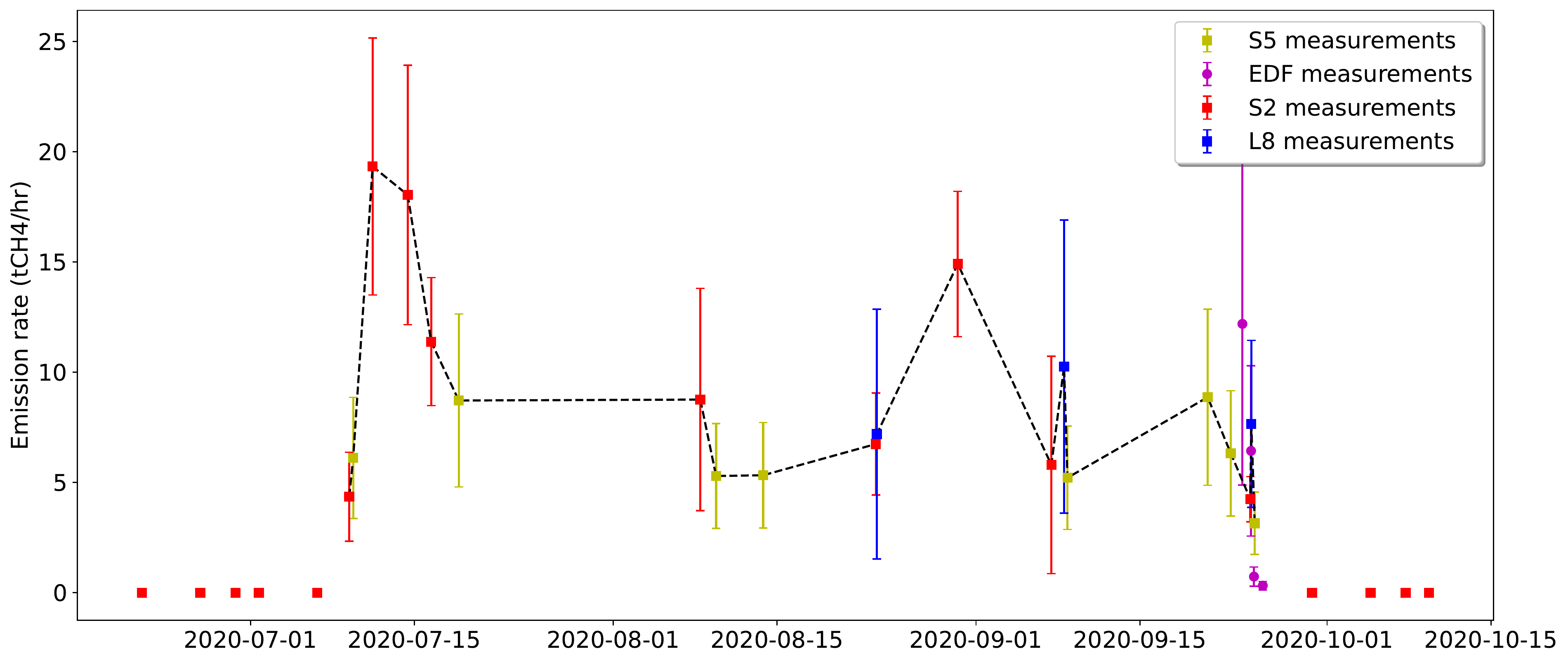}
\caption{Emission rates measured during an event in the Permian basin occurred during the summer of 2020 (estimated latitude and longitude: 31.7335, -102.0421). The plot combines estimates obtained from \emph{Sentinel-2}, \emph{Landsat-8}, \emph{Sentinel-5P} and from Scientific Aviation flights (with their respective uncertainty corresponding to the standard deviation of the estimates). All of these estimates show that the emission started more than two months earlier than it was initially reported by the EDF campaign. Moreover, these estimates seems to be consistent across sources.}
    \label{fig:permian_event}
\end{figure}

\begin{figure}
    \centering
    \includegraphics[width=\linewidth]{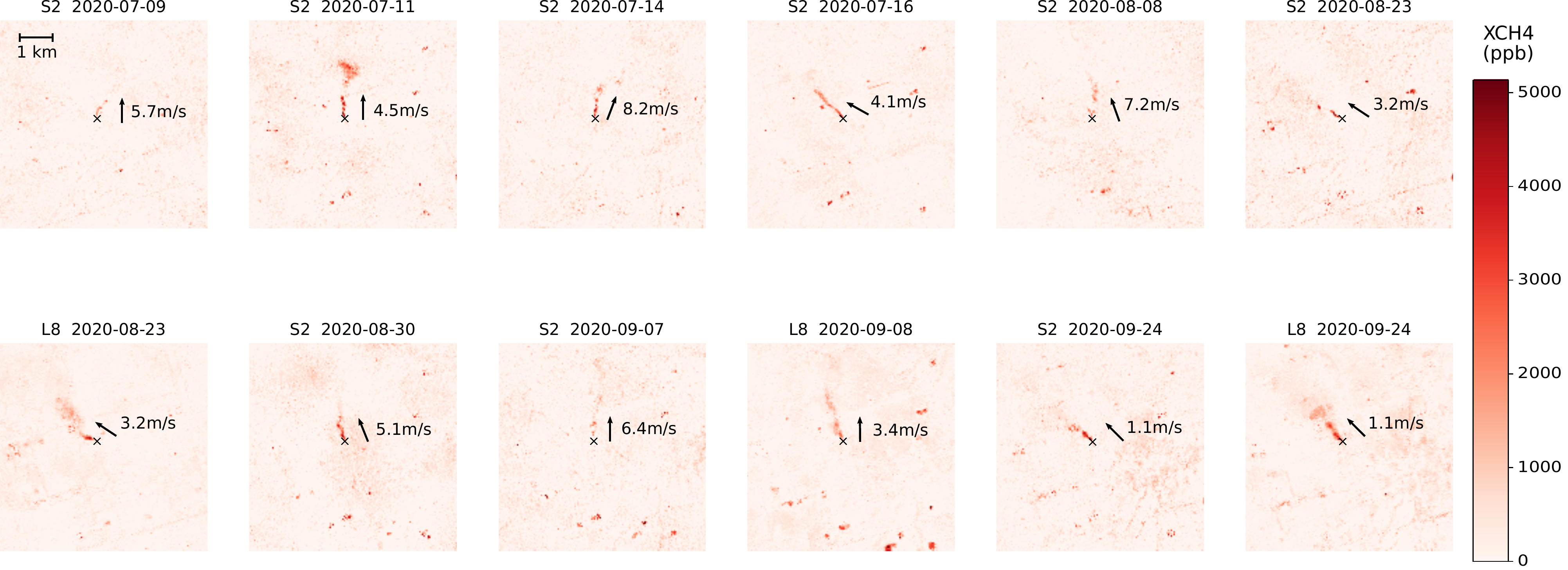} 
    \caption{Illustration of the plumes (in ppb) observed using \emph{Sentinel-2} (S2) and \emph{Landsat-8} (L8) imagery during the event in the Permian basin occurred during the summer of 2020 (estimated latitude and longitude: 31.7335, -102.0421) studied in Fig.~\ref{fig:permian_event}. The source of the emission is located at the center of the images (symbolized by the black cross) and the wind information is represented by the arrow.}
    \label{fig:permian_event_images}
\end{figure}

Fig.~\ref{fig:permian_event} shows the estimated emission rates from the mentioned sources. As we can see, the emission measurements of the airborne campaign and all those obtained after September 15$^{\text{th}}$ 2020 seems to be close and consistent with each other up until the last two EDF measurements. Yet, the analysis of the time series leads to conclude that the event had started two months prior to the aircraft campaign, thus increasing significantly the total amount of CH$_4$ released. Note that the measurements as well as the estimated confidence intervals are consistent. 
Since no emissions were detected before July 9$^{\text{th}}$ 2020 or after September 29$^{\text{th}}$ 2020, these measurements enable a full description of the event from start to end. Given that we observed a methane plume at each of the 12 dates where \emph{Sentinel-2} and \emph{Landsat-8} images were available (the plumes are shown in Fig.~\ref{fig:permian_event_images}), we assume that this event is continuous. Indeed, if that source was intermittent, we would have observed random peaks before or after our period of interest or dates with no emissions. We can therefore interpolate linearly the emission rate at a date with no image available using the two closest emission rates available. This leads to an estimation of a grand total of 16,537\textpm 7,146 tons of methane emitted during this event.

\section{Results}
\label{sec:experiments}

\begin{figure}[t]
    \centering
    \begin{subfigure}{0.32\linewidth}
        \includegraphics[width=\linewidth]{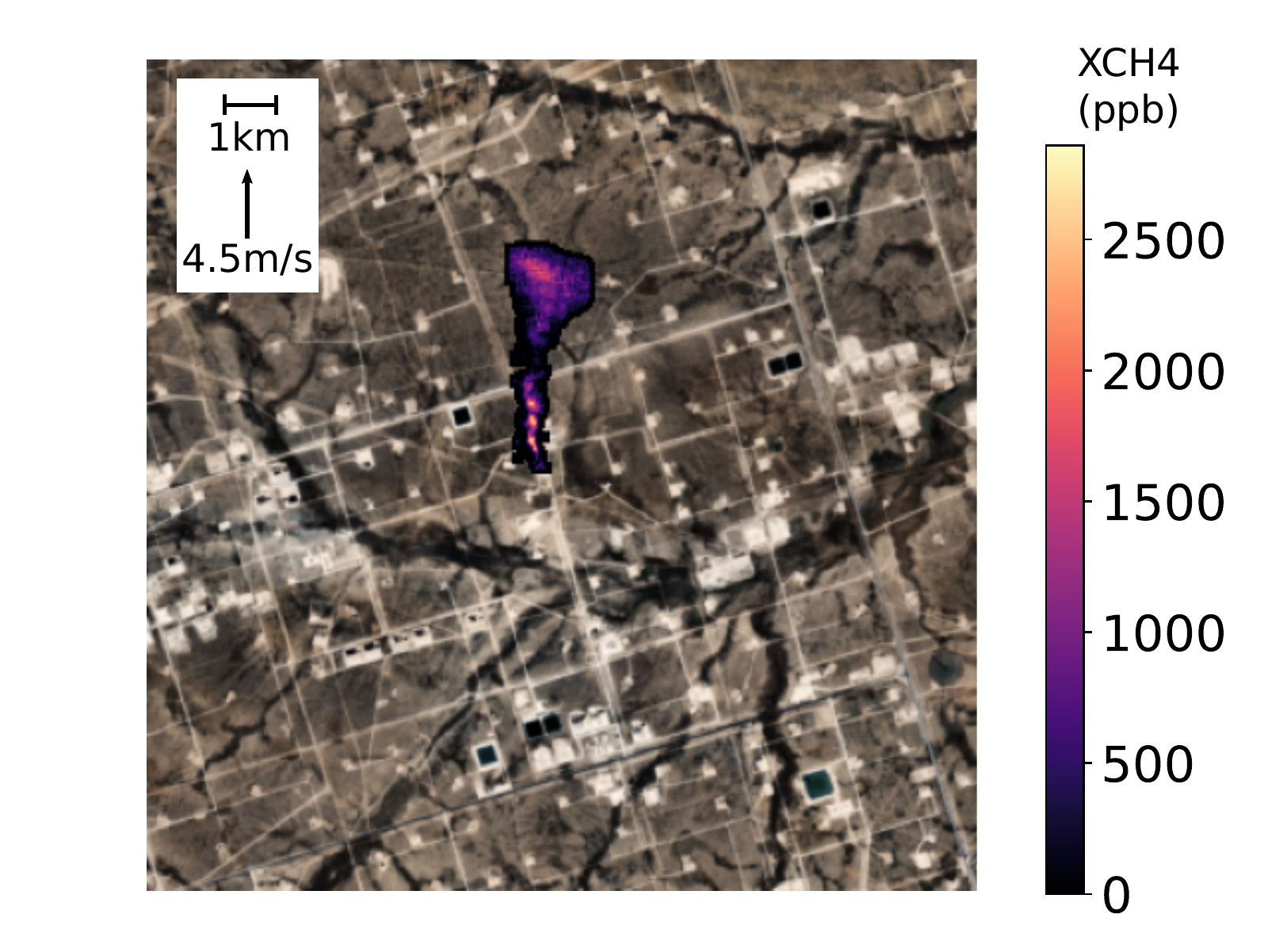}
        \caption{US, 14.5 t$_{\text{CH}_4}$/hr}
    \end{subfigure}
    \begin{subfigure}{0.32\linewidth}
        \includegraphics[width=\linewidth]{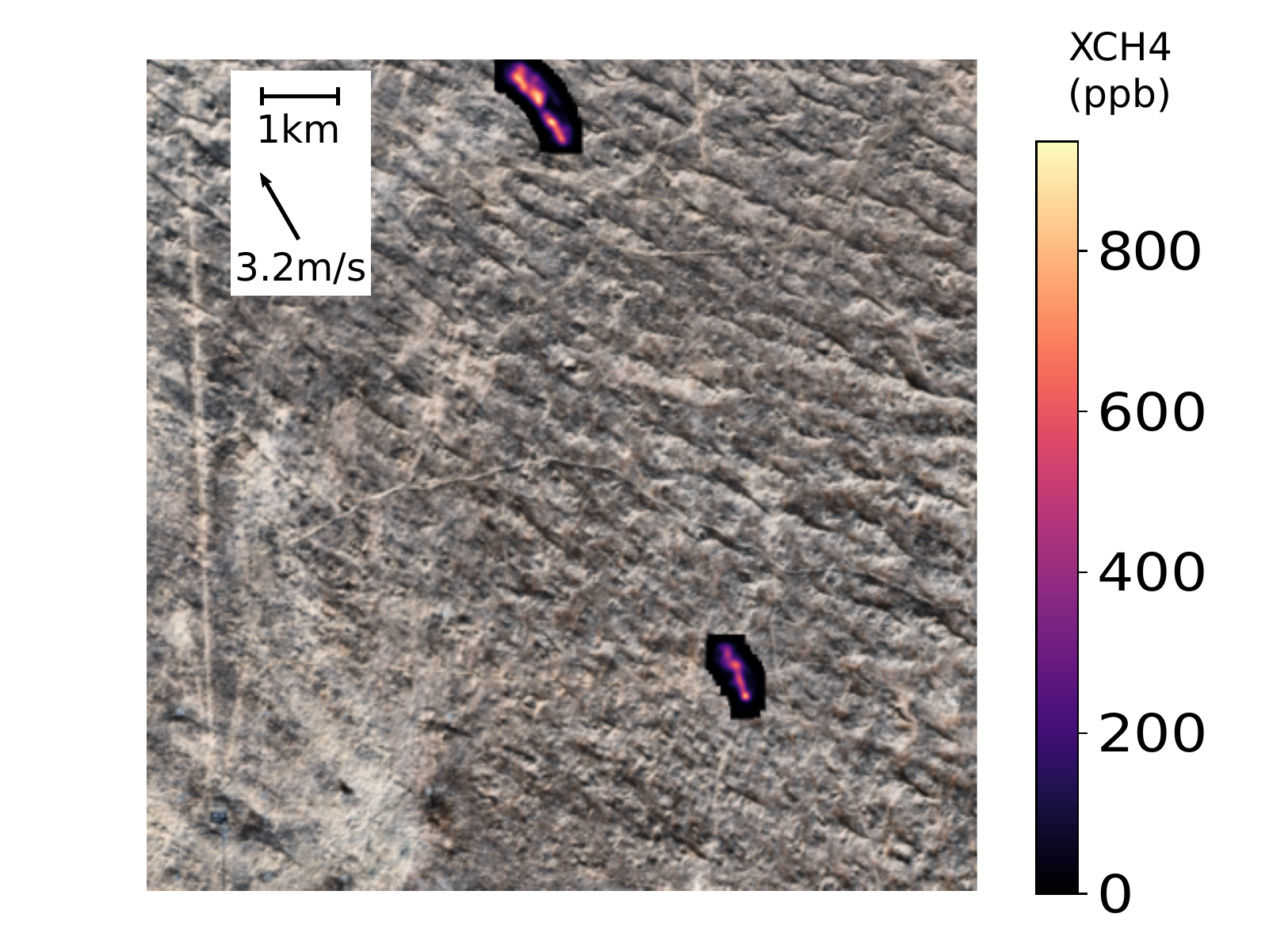}
        \caption{Turkmenistan, 8.7 t$_{\text{CH}_4}$/hr}
    \end{subfigure}
    \begin{subfigure}{0.32\linewidth}
        \includegraphics[width=\linewidth]{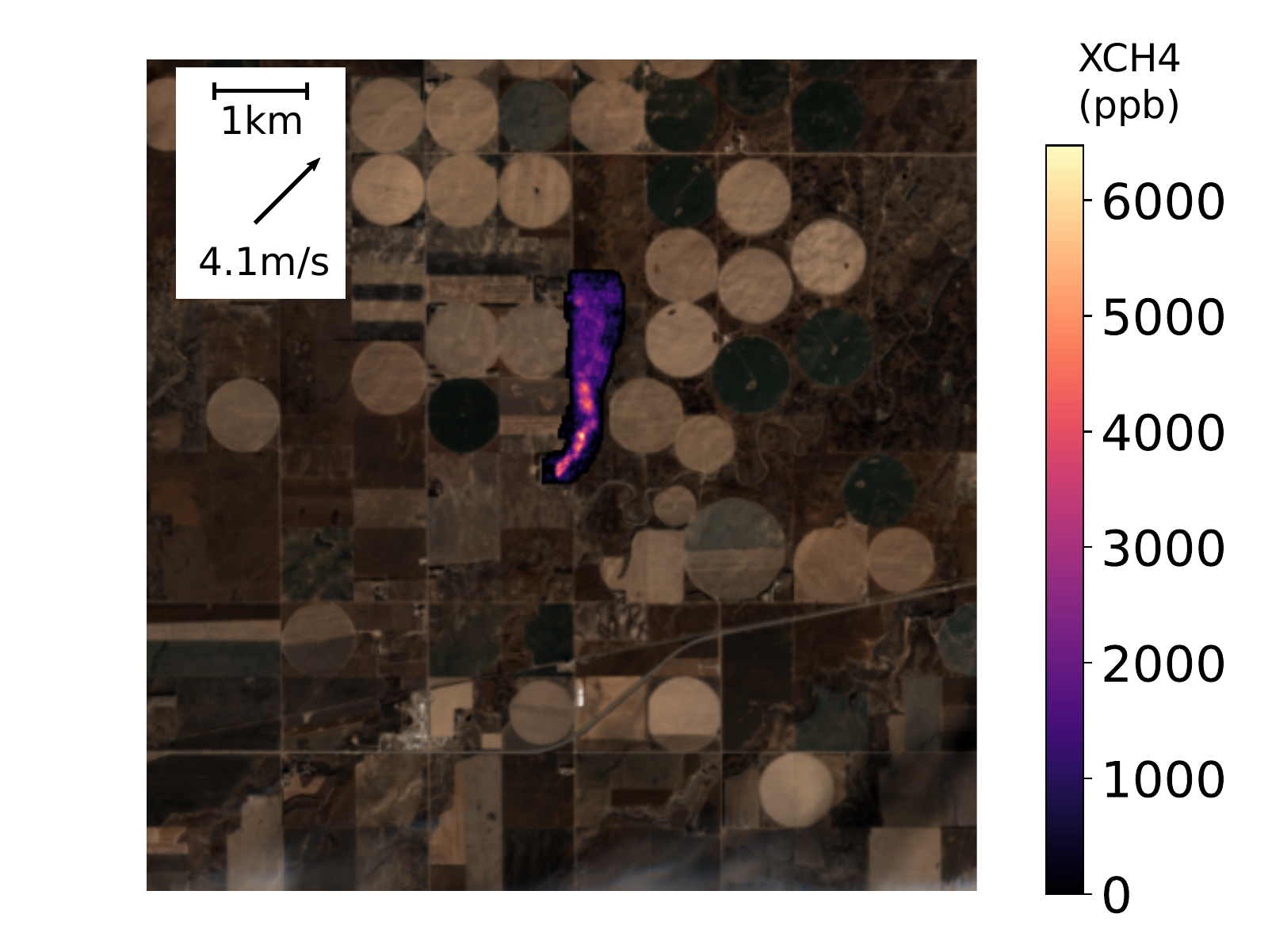}
        \caption{US, 54 t$_{\text{CH}_4}$/hr}
    \end{subfigure}
    \begin{subfigure}{0.32\linewidth}
        \includegraphics[width=\linewidth]{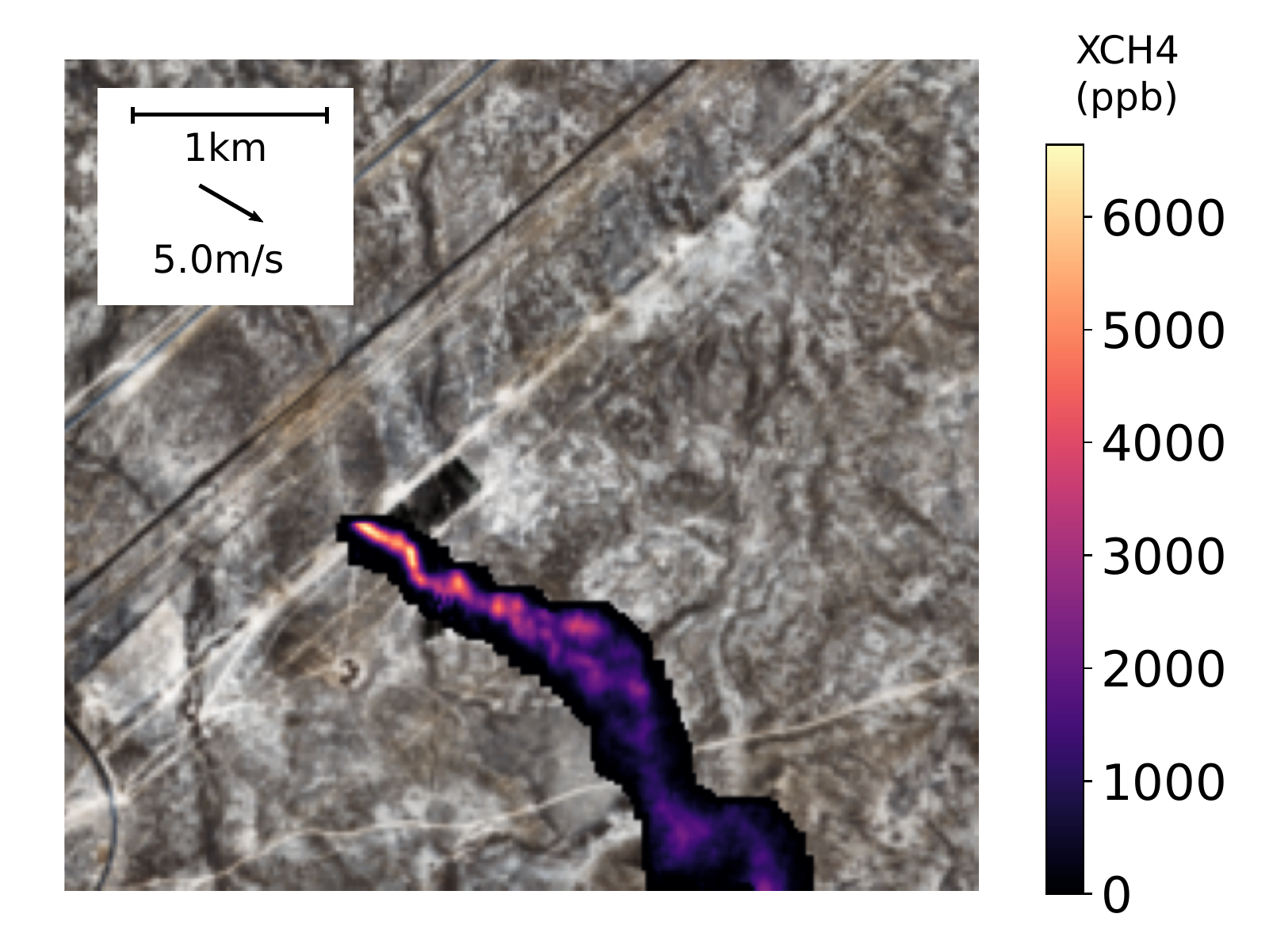}
        \caption{Kazakhstan, 62.8 t$_{\text{CH}_4}$/hr}
    \end{subfigure}
    \begin{subfigure}{0.32\linewidth}
        \includegraphics[width=\linewidth]{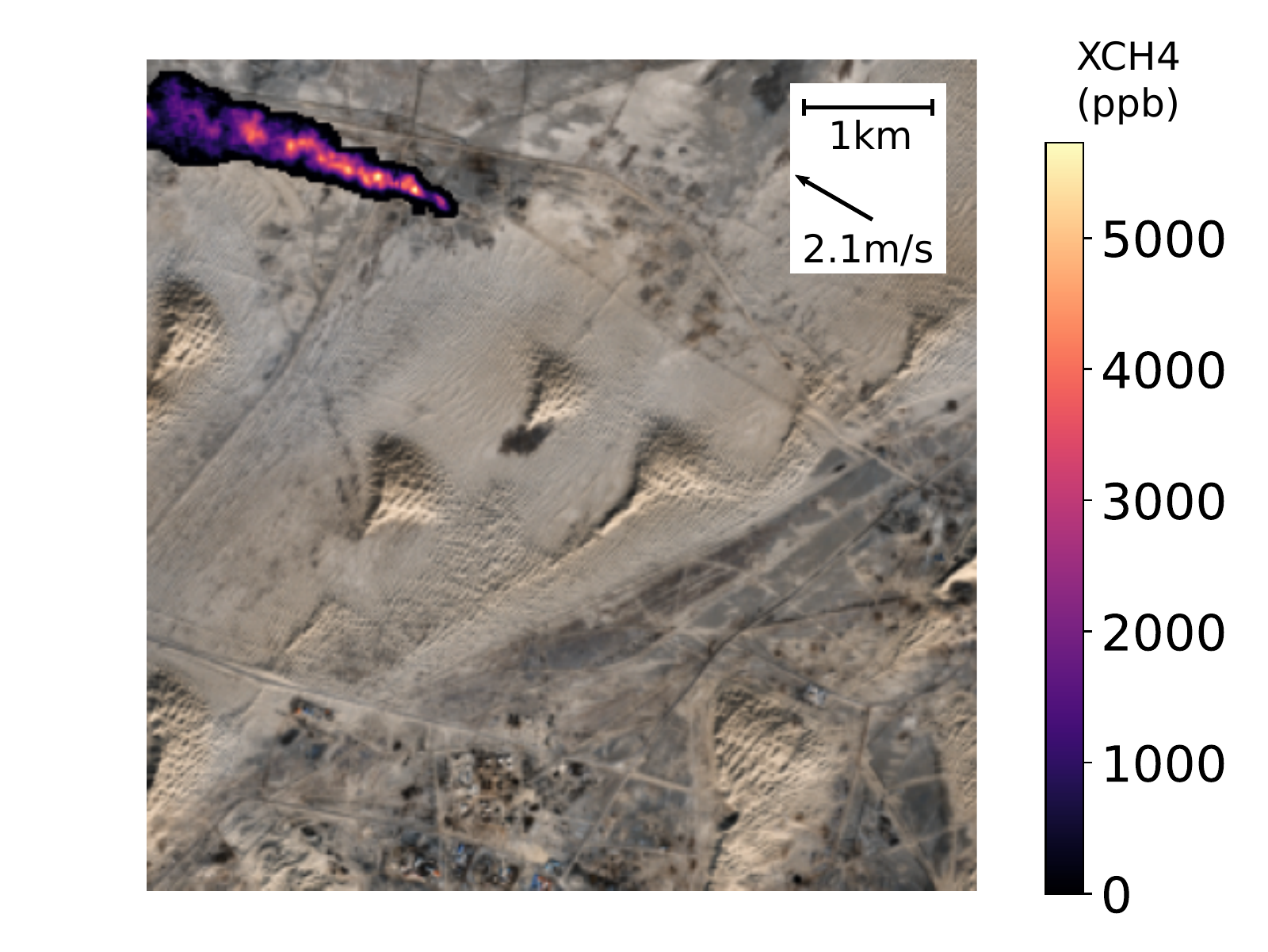}
        \caption{Turkmenistan, 41 t$_{\text{CH}_4}$/hr}
    \end{subfigure}
    \begin{subfigure}{0.32\linewidth}
        \includegraphics[width=\linewidth]{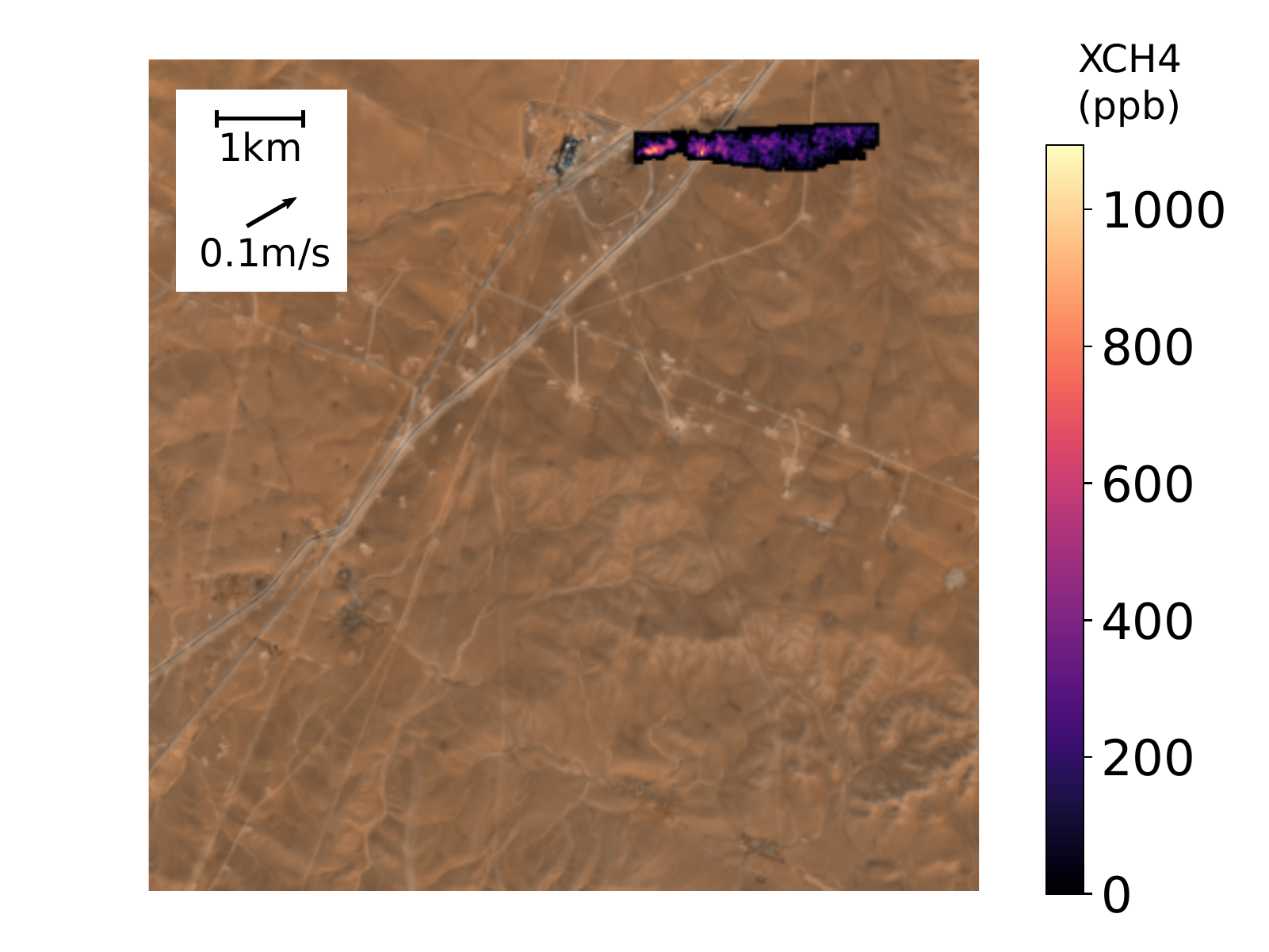}
        \caption{Algeria, 3.1 t$_{\text{CH}_4}$/hr}
    \end{subfigure}
    \begin{subfigure}{0.32\linewidth}
        \includegraphics[width=\linewidth]{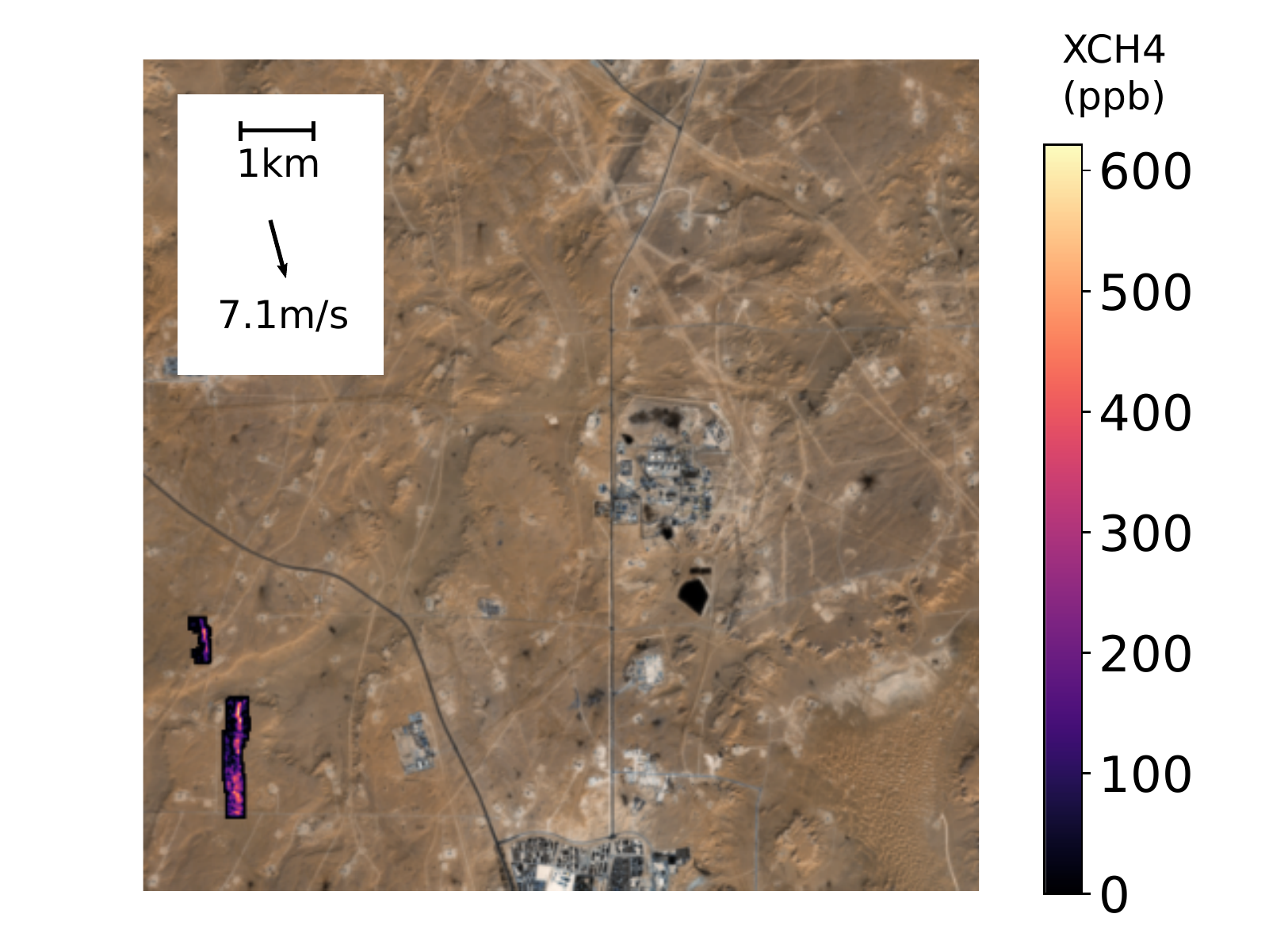}
        \caption{Algeria, 5.7 t$_{\text{CH}_4}$/hr}
    \end{subfigure}
    \begin{subfigure}{0.32\linewidth}
        \includegraphics[width=\linewidth]{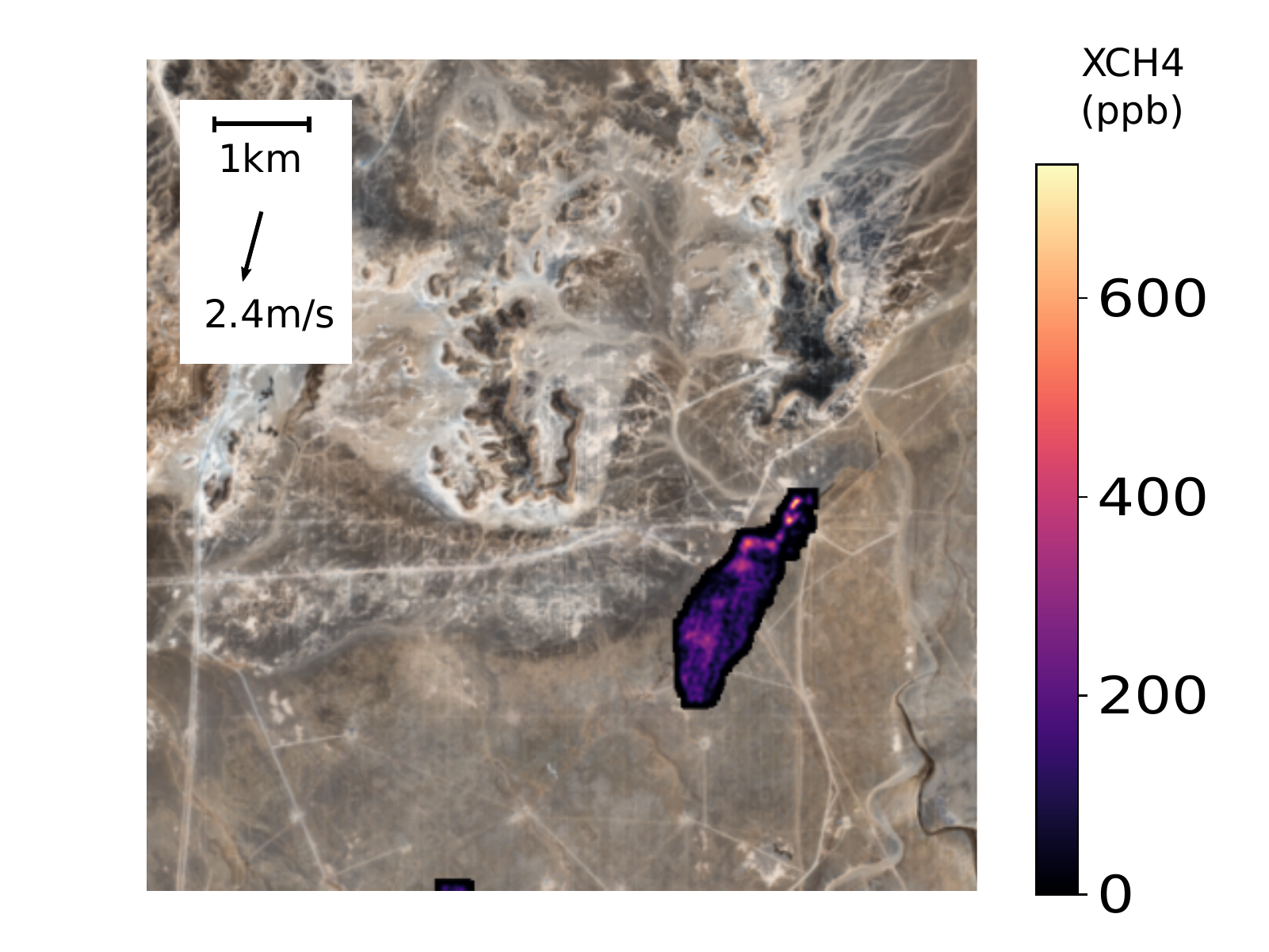}
        \caption{Algeria, 3.3 t$_{\text{CH}_4}$/hr}
    \end{subfigure}
    \begin{subfigure}{0.32\linewidth}
        \includegraphics[width=\linewidth]{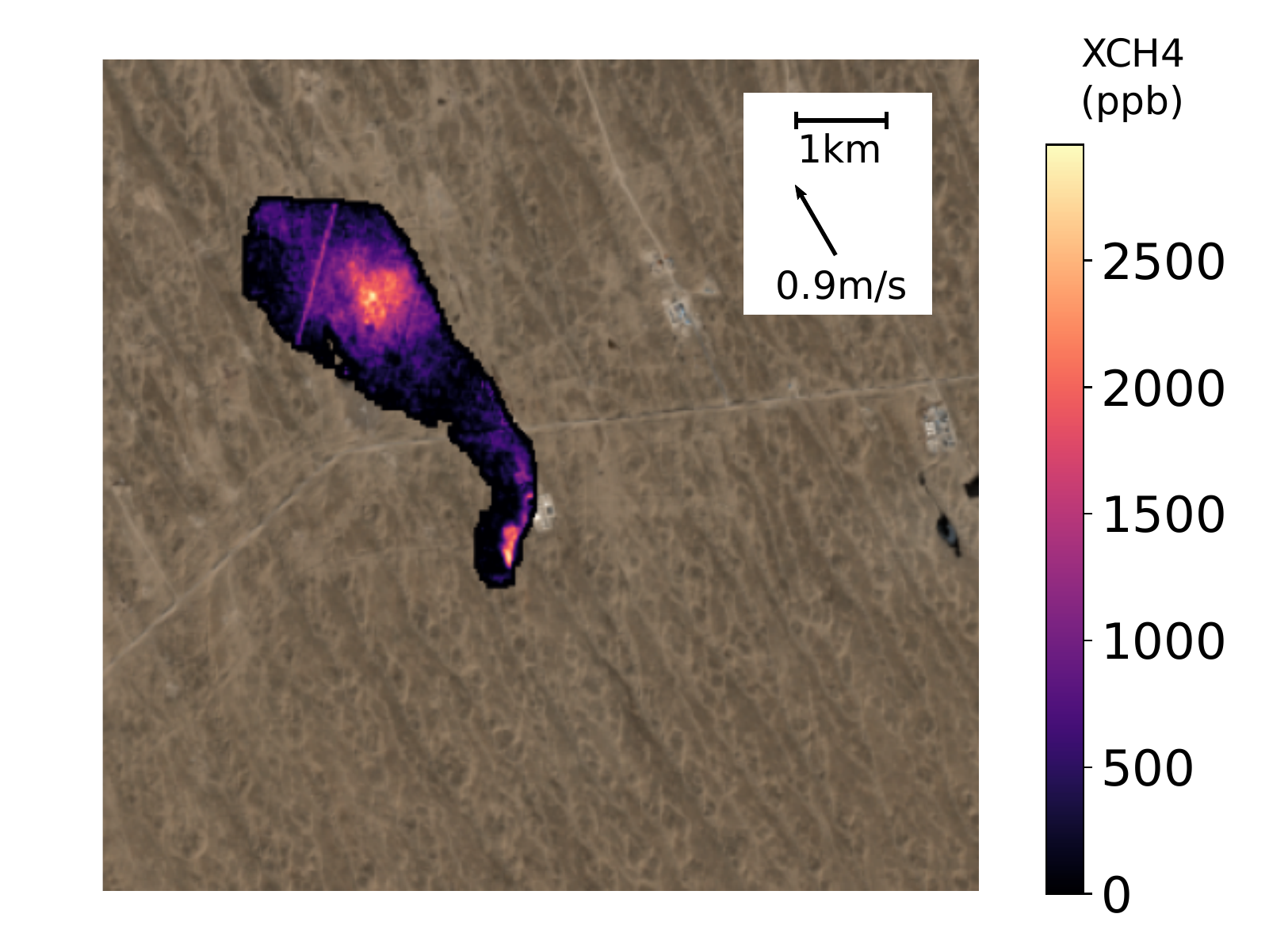}
        \caption{Turkmenistan, 27.5 t$_{\text{CH}_4}$/hr}
    \end{subfigure}
    \caption{Examples of detected plumes over oil and gas facilities. The plume is shown as a white and purple mask over the corresponding \emph{Sentinel-2} satellite RGB images (top of atmosphere). Wind information is represented by the arrow.}
    \label{fig:plumes_example}
\end{figure}

\begin{figure}
    \centering
    \sbox{\bigpicturebox}{%
        \begin{subfigure}[b]{.50\textwidth}
            \includegraphics[width=\textwidth]{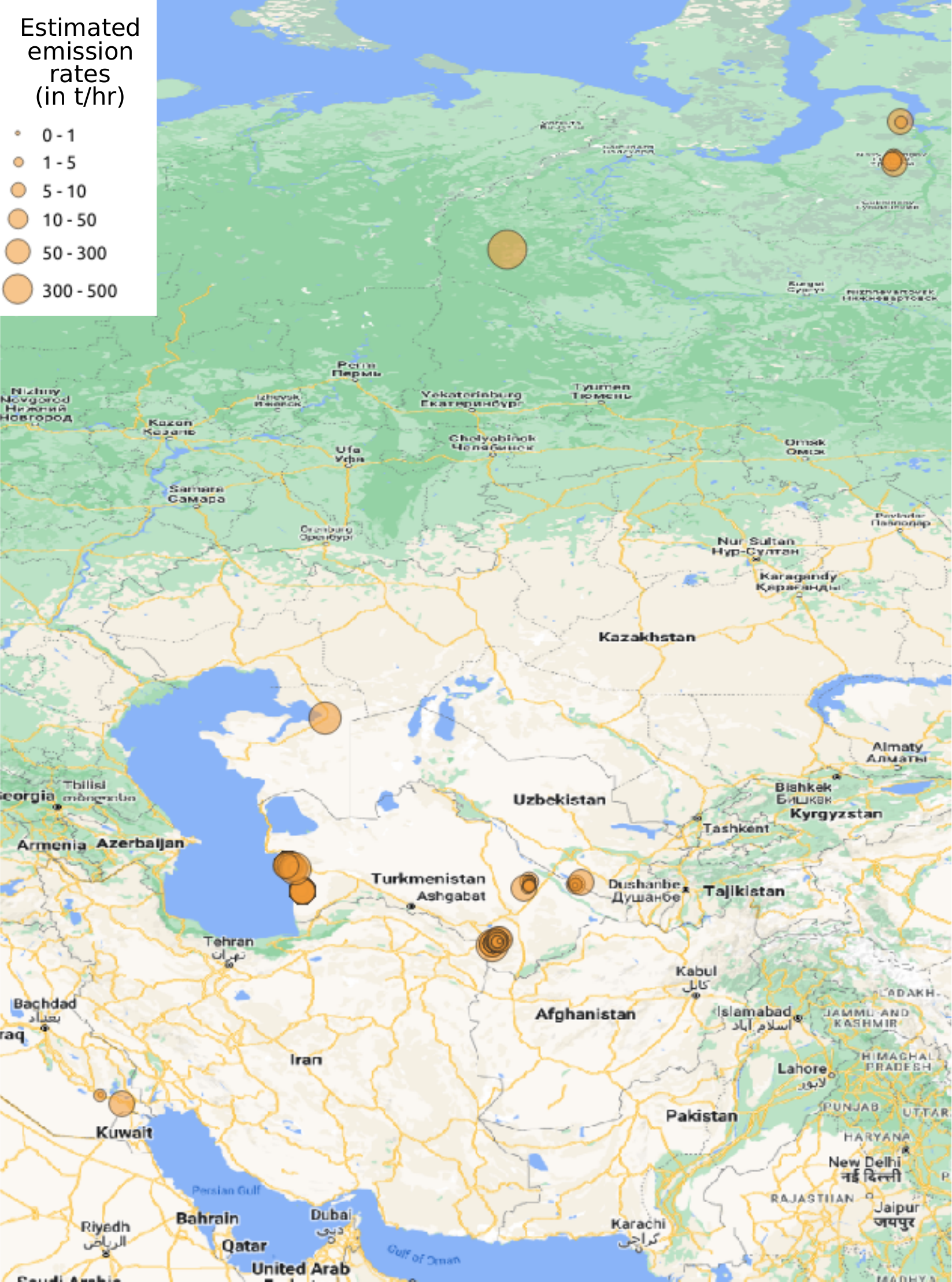}%
            \caption{Location of the plumes detected in the Turkmenistan region.}
        \end{subfigure}
    }

    \usebox{\bigpicturebox}
    \begin{minipage}[b][\ht\bigpicturebox][s]{.45\textwidth}
        \begin{subfigure}{\textwidth}
            \includegraphics[width=\textwidth]{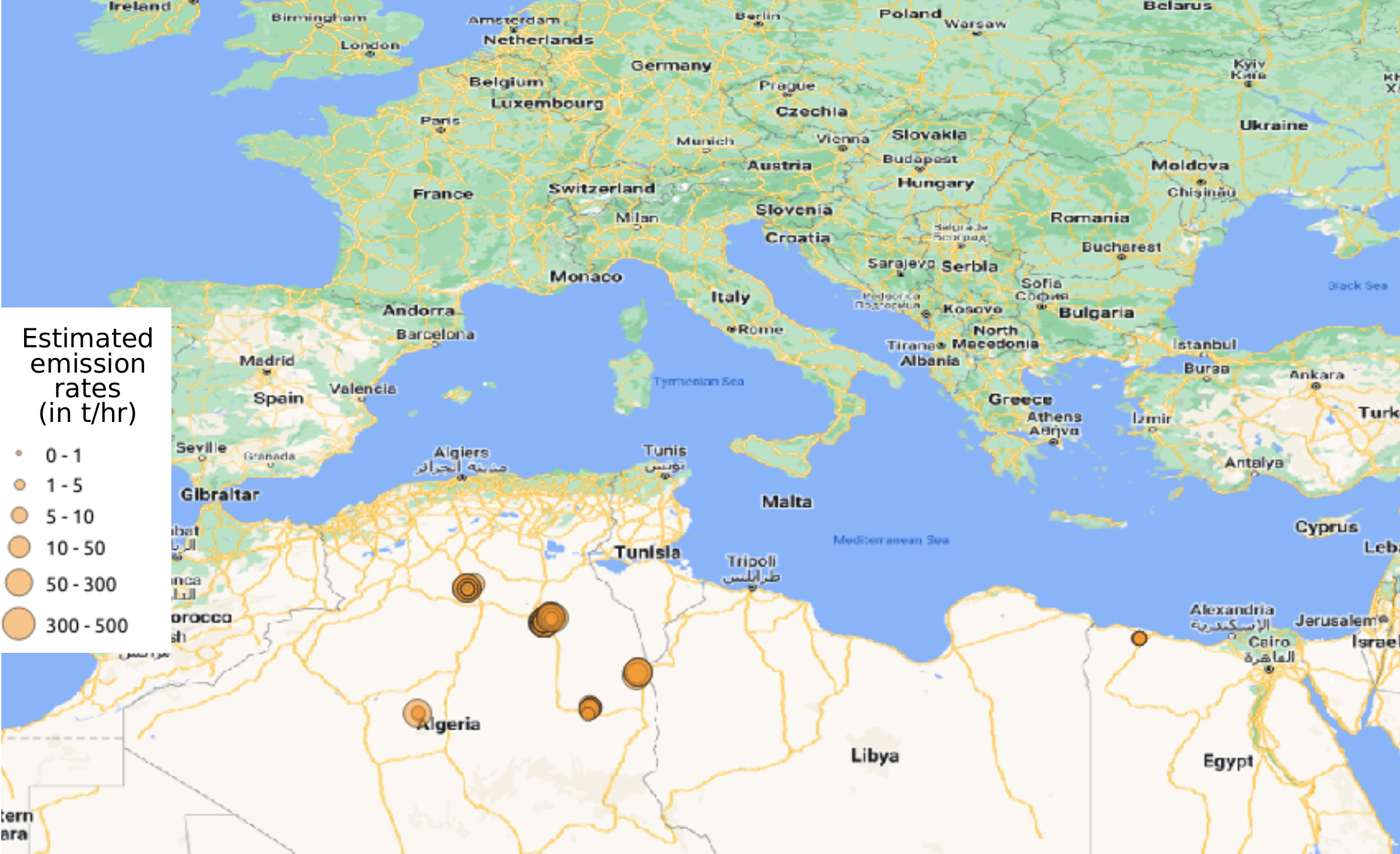}
            \caption{Location of the plumes detected in the Algeria region.}
        \end{subfigure}\hfill
        \begin{subfigure}{\textwidth}
            \includegraphics[width=\textwidth]{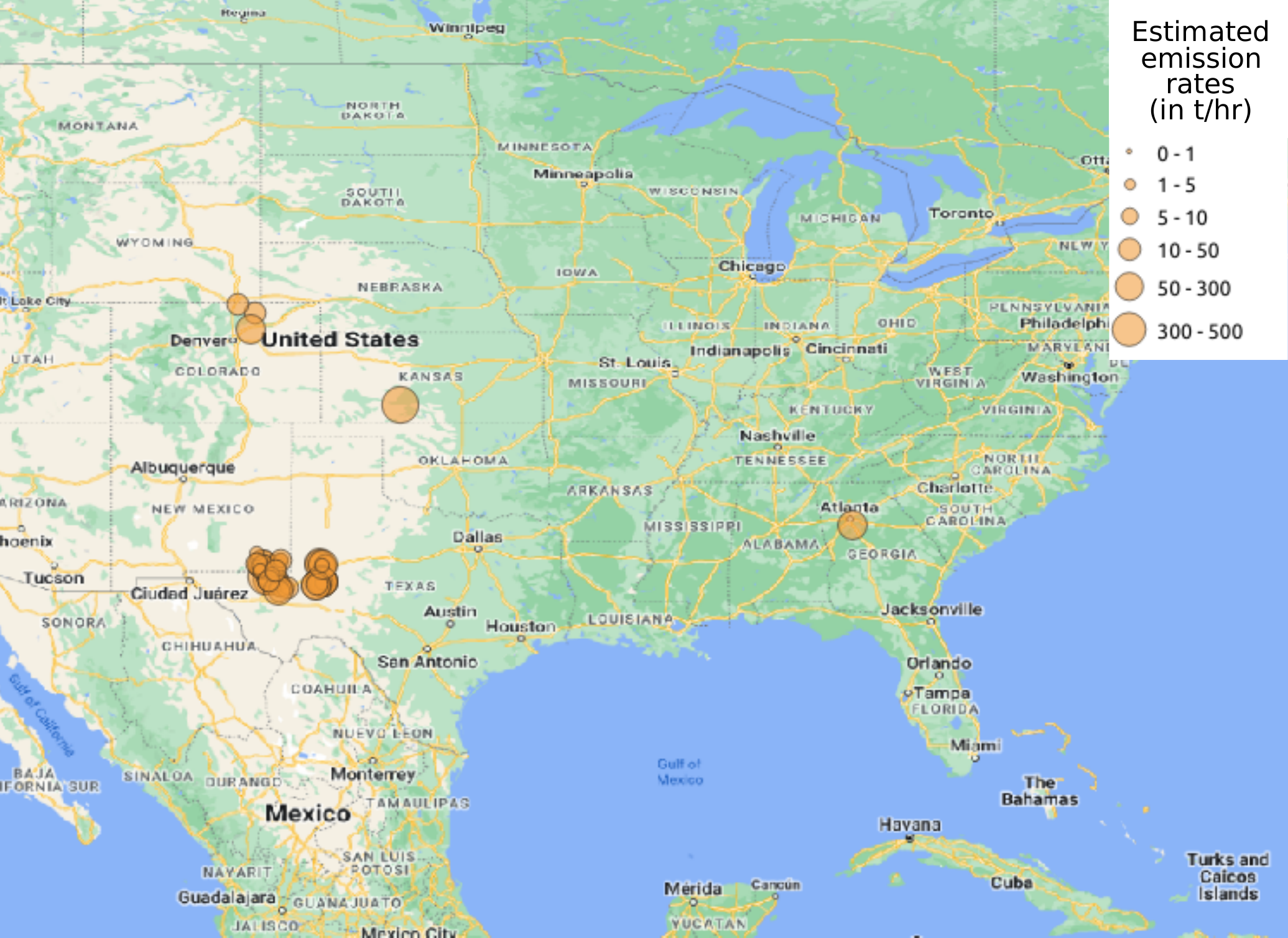}
            \caption{Location of the plumes detected in the US.}
        \end{subfigure}
    \end{minipage}
    \vspace{-7em}

    \caption{Locations of the different detected emissions corresponding to the proposed dataset. The quantified emission rates are illustrated by the circles' diameter.}
    \label{fig:map_of_leaks} 
\end{figure}

We monitored about 7000 geographical sites of interest linked to oil and gas facilities during a period of 47 months, from November 2017 to September 2021. Every site is associated to a 10x10km tile. For every tile, a time series of at least six months was extracted. In total for this study, more than 1248621 (potentially cloudy) tiles were processed  over 562652 km$^2$. The proposed dataset comprises a location and a quantification for all plumes. Each detected plume in the dataset is quantified using the IME method (see the \emph{Source quantification} subsection). Fig.~\ref{fig:plumes_example} shows a selection of methane plumes from the proposed dataset. We also associate to each plume the corresponding wind data from ECMWF-ERA5~\cite{ERA5} that is used to quantify the emission.%
As of September 2021, 1202 plumes were detected using \emph{Sentinel-2} images from 92 different sites of interest, mostly located in three countries: Algeria, Turkmenistan and the United States (see Fig.~\ref{fig:map_of_leaks}). 
Table~\ref{tab:detections_by_countries} shows the number of detected events per country. Note that the imbalance of detection in the US compared to the Algeria in the proposed dataset is mostly due to more difficult sensing conditions. Indeed, it was shown by Gorroño~\etal~\cite{gorrono2021} through simulations that the detection limit of \emph{Sentinel-2} can be expected to be in the 8 to 12 t$_{\text{CH}_4}$/hr in the Permian basin in the US while it should be in the 1.5 to 2.5 t$_{\text{CH}_4}$/hr range in Turkmenistan. We empirically verified these expected detection limits through our detections with \emph{Sentinel-2}.

\begin{figure}[t]
    \centering
    \includegraphics[width=.6\linewidth]{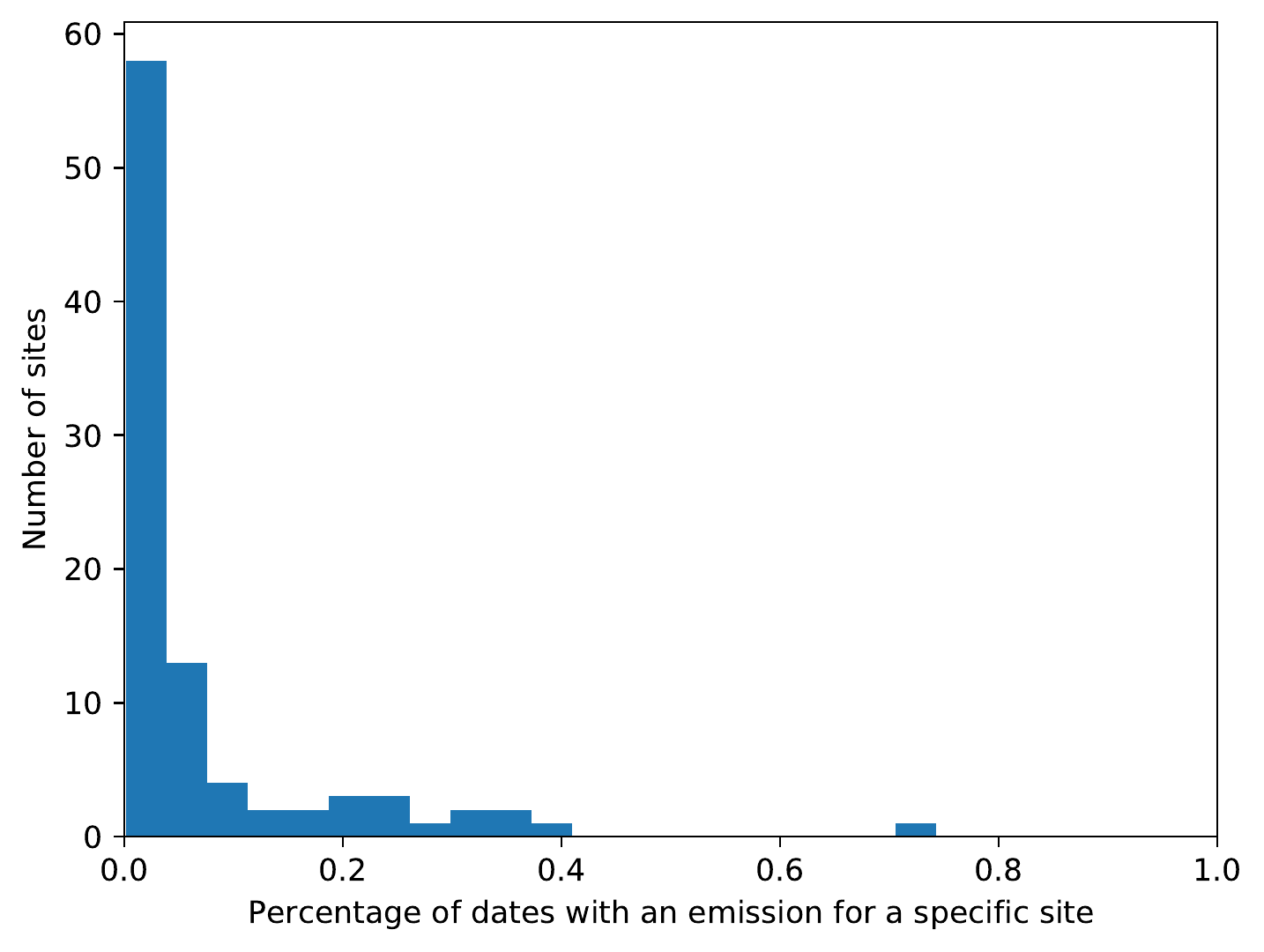}
    \caption{Histogram of recurrence of emissions.}
    \label{fig:stats_plumes}
\end{figure}

We then classified these emissions into two main categories: recurrent and unique. An event is said to be recurrent when at least two methane plumes have been detected in the time series of a given area of interest. The rest of the emission are characterized as unique \ie only one plume was detected in the considered area of interest in the entire time series. We found that $58\%$ of these plumes could be attributed to recurrent events. This means that these events are likely not due to an unexpected major incident, and could probably be avoided with better monitoring and maintenance of oil and gas facilities. We present a more detailed histogram of recurrence of emissions in Fig.~\ref{fig:stats_plumes}.

\begin{table}
\small
    \centering
    \begin{tabular}{l|c}
        \bf Country  & \bf Number of events  \\
        \hline
        Algeria (DZA) & 527 \\
        Turkmenistan (TKM) & 526 \\
        United States (USA) & 98 \\
        Uzbekistan (UZB) & 27 \\
        Egypt (EGY) & 13 \\
        Russian Federation (RUS) & 7  \\ 
        Iraq (IRQ) & 3 \\
        Kazakhstan (KAZ) & 1\\
        \hline
    \end{tabular}
    \caption{Distribution by country of the detected emissions during the period of 47 months going from November 2017 to September 2021.}
    \label{tab:detections_by_countries}
\end{table}

\subsection{Global power law fitting}
\label{sec:powlaw_global}

\begin{figure}
    \centering
    \begin{subfigure}{.8\linewidth}
        \includegraphics[width=\linewidth]{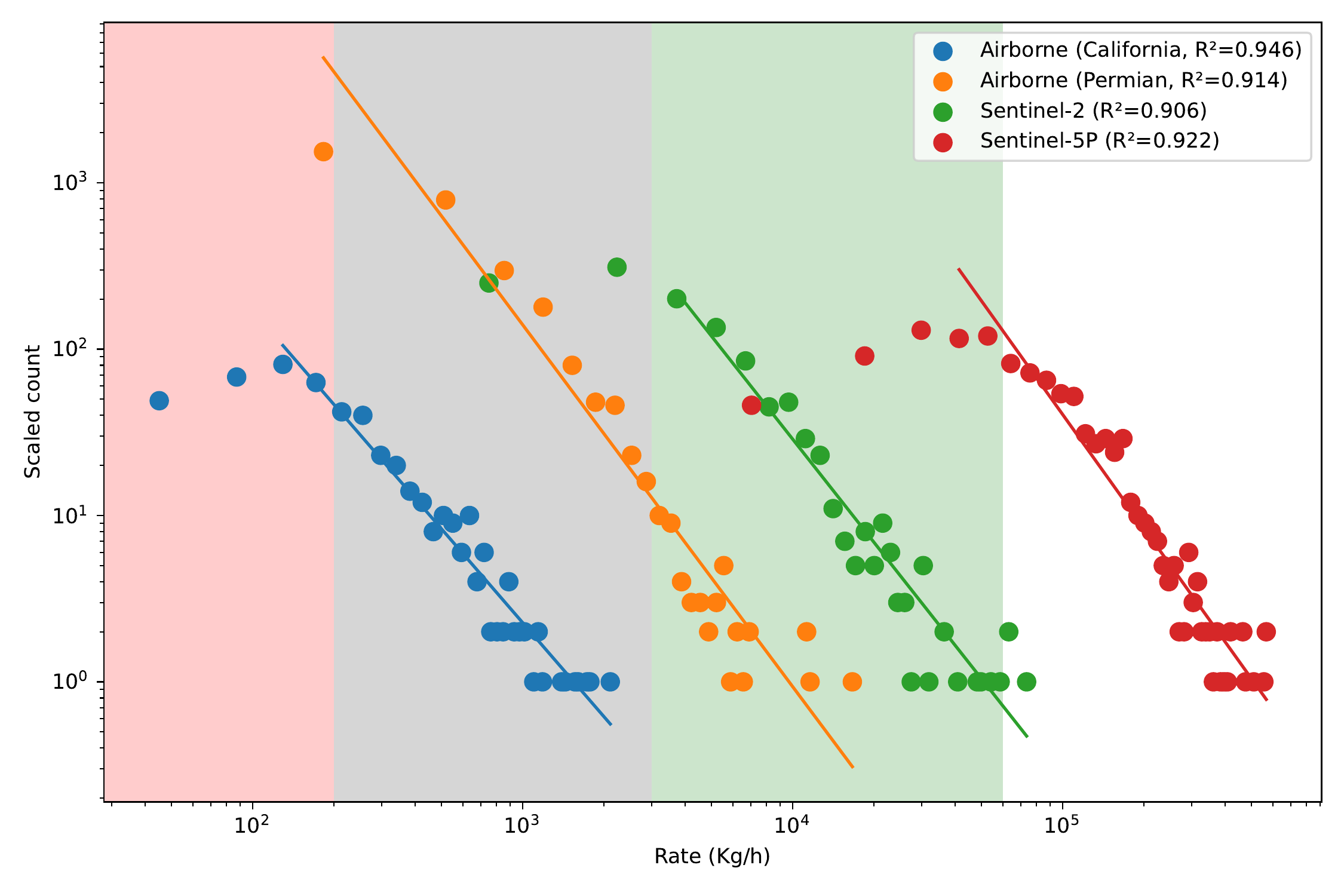}
        \caption{Before scaling}
    \end{subfigure}
    \begin{subfigure}{.8\linewidth}
        \includegraphics[width=\linewidth]{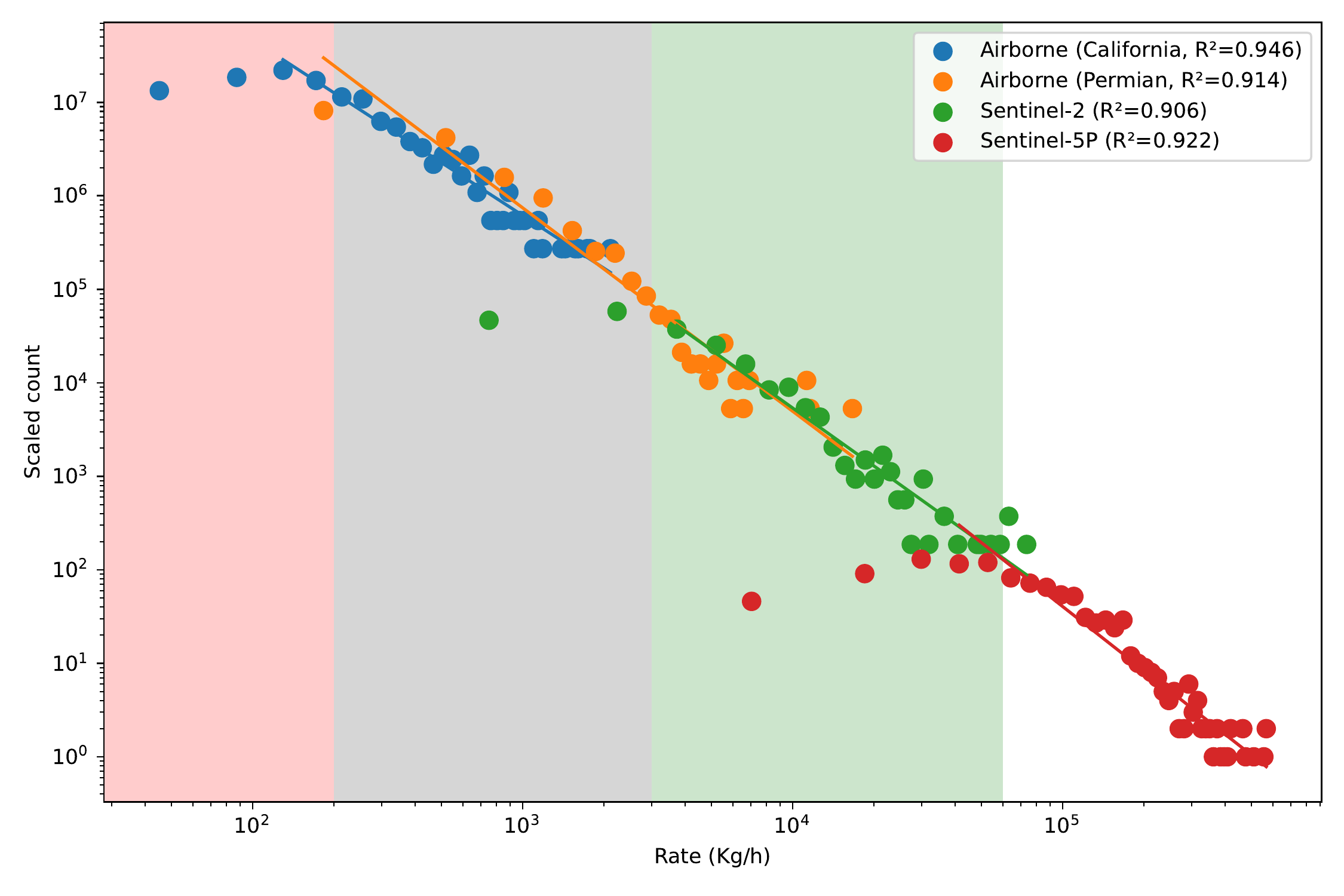}
        \caption{After scaling}
    \end{subfigure}
    \caption{Power law plot of \emph{Sentinel-5P} and \emph{Sentinel-2} events, together with airborne campaigns over California~\cite{duren2019california} and the Permian~\cite{cusworth2021intermittency}. Counts are scaled to match in common detection zones. Since data is consistent across the different sources used for the study, this shows that at a global scale large event observation might be a good proxy indicator for smaller but unobserved events (corresponding to the green region). The grey region represent the region where the model still be valid. It is however unlikely that the power law model is still valid in the red region.}
    \label{fig:power_law}
\end{figure}

Power law models have been shown to be a good fit for many GHG emission studies~\cite{akhundjanov2017size,marzadri2020power,elder2020airborne}. In a recent paper, Lauvaux~\etal~\cite{lauvaux2021global} postulated that the methane emission events follow a power law distribution at a global scale. This was observed using emission rates estimated from \emph{Sentinel-5P} and airborne hyperspectral measurements.
In this work, we merged the events from our \emph{Sentinel-2} based dataset into the previously proposed power law plot~\cite{lauvaux2021global} to complete the picture. The power law that we obtained is shown in Fig.~\ref{fig:power_law}. 
We rescaled counts for \emph{Sentinel-2} and airborne campaigns so that counts match for all sources for emissions rates where events can be detected by multiple sources. 
The rationale behind this scaling is that, everything else being equal, detection counts should match for all sources for which emission rates are above the detection limit. The scaling is thus meant to compensate for differences in spatial coverage, revisit frequency, weather impact, etc. This is justified by the study of the Permian event where we show that given the proper sensing conditions detections match across the different source type considered here.
In practice, this means that \emph{Sentinel-2} counts are scaled to match \emph{Sentinel-5P} counts at 50 t$_{\text{CH}_4}$/hr, while California~\cite{duren2019california} and Permian~\cite{cusworth2021intermittency} airborne campaigns counts are scaled to match S2 counts at 5 t$_{\text{CH}_4}$/hr. For that, the regressions are first estimated for each source independently and the scaling is then done based on the estimated regression by changing the intercept to ones that align the different models.
For example, consider that the detections from \emph{Sentinel-5P} can be modeled using $\beta_{S5}x^{\alpha_{S5}}$ and the detections from \emph{Sentinel-2} by $\beta_{S2}x^{\alpha_{S2}}$. In that case, the scaling for \emph{Sentinel-2} corresponds to $\beta_{S2,scale} = \beta_{S5} x_{S2,S5}^{\alpha_{S5} - \alpha_{S2}}$ where $x_{S2,S5} = 50 t_{\text{CH}_4}/hr$ as mentioned previously. This means that the scaled model for \emph{Sentinel-2} coincides with the model for \emph{Sentinel-5P} in $x_{S2,S5} = 50 t_{\text{CH}_4}/hr$ but the slope of the model is not changed.

We define the detection limit as the threshold that represents the regime in which, except in the most adverse conditions, sources should be detected. In practice, it corresponds to the point below which the linear models is not valid anymore since detections are missed. This phenomenon is visible in Fig.~\ref{fig:power_law} where each curve ``tails off" on the lower end.
This also means that it is possible to detect emissions smaller than this limit when conditions are optimal (e.g appropriate wind conditions, good atmospheric conditions and good surface reflectance). In order to have more robust models, the estimation of the power laws is done using only the data above their detection limits.

It also seems that there is a maximum detection limit for \emph{Sentinel-2}. We think that it is just because, in practice, these events are rare. \emph{Sentinel-5P} is capable of detecting very small excesses of methane  -- over a large spatial region -- thanks to its hyperspectral sensor. As such, it is  able to detect a plume even -- closely -- after the end of an event. On top of that it provides a large spatial coverage and better revisit frequency. In the end, it is much more likely to find ultra-emitters with \emph{Sentinel-2} than \emph{Sentinel-5P} and this explains the apparent maximum detection limit in Fig.~\ref{fig:power_law}. It is also possible that some of these events have been under-quantified because of the difficulty to annotate very large events (similarly to how it is difficult to quantify properly ultra-emitters with airborne campaigns).

Remark that \emph{Sentinel-2} observations are well aligned with the \emph{Sentinel-5P} power law slope and complete the range for medium scale events, bridging the gap in emission rates between small sources (0.1 t$_{\text{CH}_4}$/hr to 10 t$_{\text{CH}_4}$/hr) captured by airborne campaigns and the ultra-emitters ($>$ 25 t$_{\text{CH}_4}$/hr) detected by, for example, \emph{Sentinel-5P}. This shows that at a global scale large event observations seen by \emph{Sentinel-5P} are a good proxy indicator for smaller events in the range covered by \emph{Sentinel-2} (\ie $>$ 25 t$_{\text{CH}_4}$/hr) showed in green in Fig.~\ref{fig:power_law}. The good alignment with the data from the airborne campaign hints that the model might still be valid for even smaller events in the range covered by these campaigns (\ie $>$ 0.1 t$_{\text{CH}_4}$/hr) showed in gray in Fig.~\ref{fig:power_law}. It is however unlikely that the power-law is still valid for even smaller events corresponding to the last region shown in red in Fig.~\ref{fig:power_law}. There exists a theoretical limit to the power law that has not been identified yet. Most likely, once observing small leaks from pressure valves, it is expected to see a major shift in the distribution.

Our results suggest that one could only monitor the largest emitters and draw conclusions on smaller emitters, assuming that the slope is defined by structural variables (for example the ratio between small and large pipes) and by the maintenance operation procedures. Assuming that the slope of the power law will remain, one could track progress by only looking at a fraction of the top emitters and extrapolate the observed trend to smaller emitters.
While the statistical relationship (Power Law) observed across our data set suggests that large emitters offer indirect monitoring of smaller emitters, the linear coefficients from such regression will be specific to each producing region. Additional data are necessary to constrain more precisely the slopes of regional Power Law relationships.

\subsection{Per country analysis}
\begin{figure}[t]
    \centering
    \begin{subfigure}{.45\linewidth}
    \includegraphics[width=\linewidth]{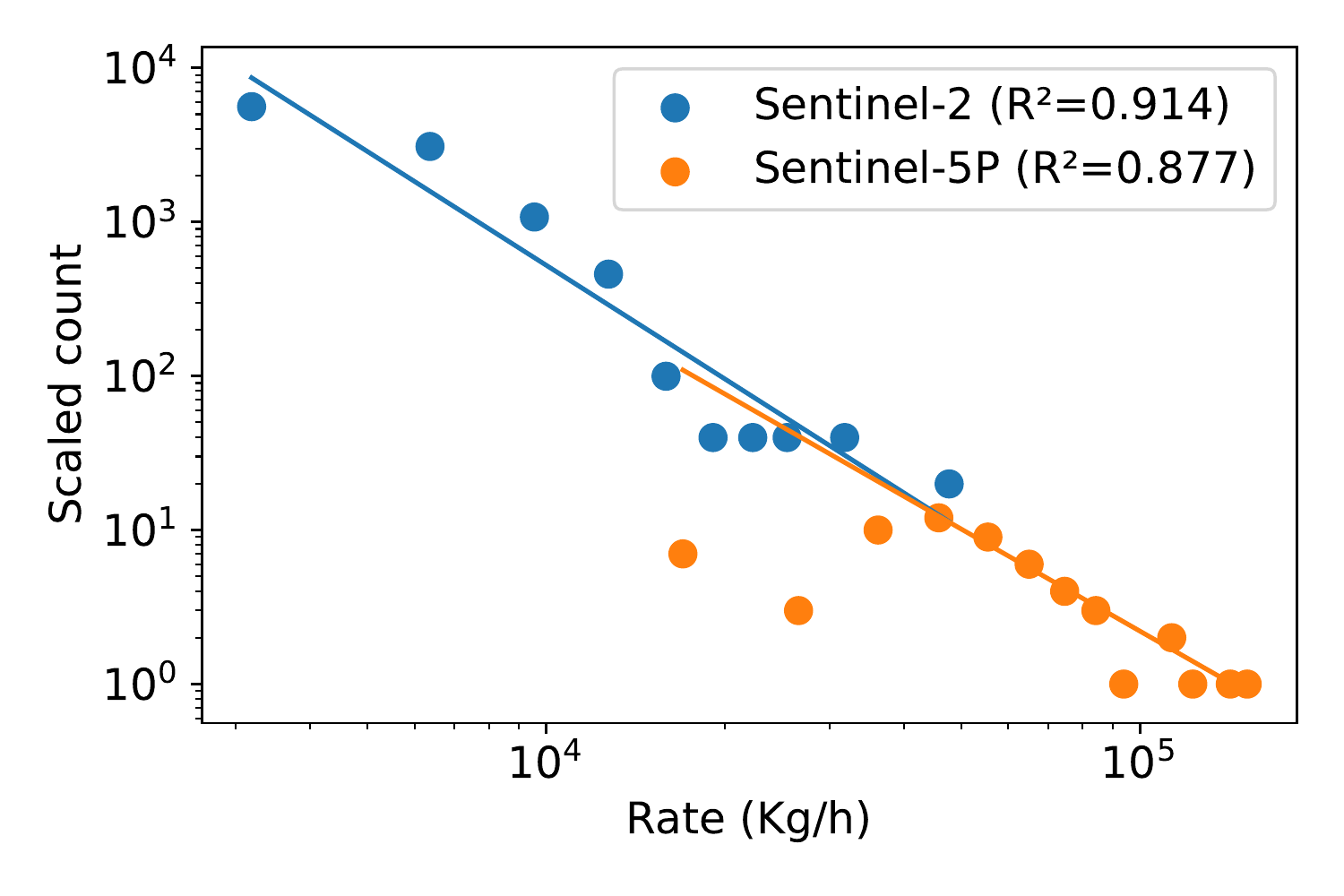}
    \caption{Local power law for the region of Algeria. \phantom{additional text}}
    \end{subfigure}
    \begin{subfigure}{.45\linewidth}
    \includegraphics[width=\linewidth]{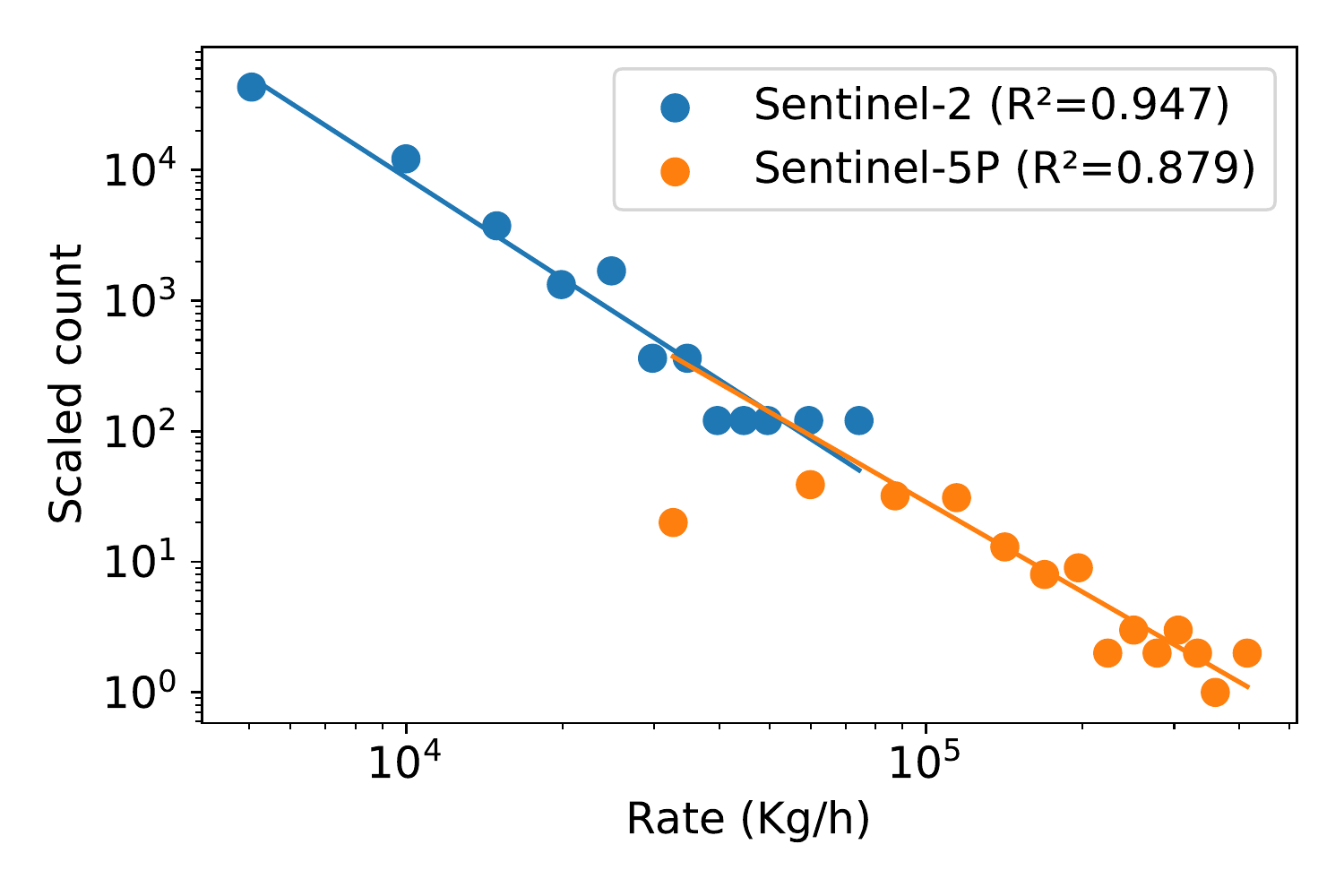}
    \caption{Local power law for the region of Turkmenistan.}
    \end{subfigure}
    \caption{Power law by country using \emph{Sentinel-2} and \emph{Sentinel-5P} measurements. This seems to indicate the power law is still valid at a regional level. Left: Algeria, right: Turkmenistan}
    \label{fig:powerlaw-country}
\end{figure}

We analyzed the previous data on a per-country basis. We considered only measurements from \emph{Sentinel-5P} and \emph{Sentinel-2} and studied the detections in Algeria and Turkmenistan. This analysis is shown in Fig.~\ref{fig:powerlaw-country}. Extending the global power law presented previously, these two countries exhibit a similar power law model at a regional level. This means that not only ultra-emitters ($>$ 25 t$_{\text{CH}_4}$/h) are a proxy to large-emitters ($>$ 2 t$_{\text{CH}_4}$/h) at a global scale, they might also be a good proxy at a more local, \eg basin, level.

\begin{figure}[t]
    \centering
    \begin{subfigure}{.45\linewidth}
    \includegraphics[width=\linewidth]{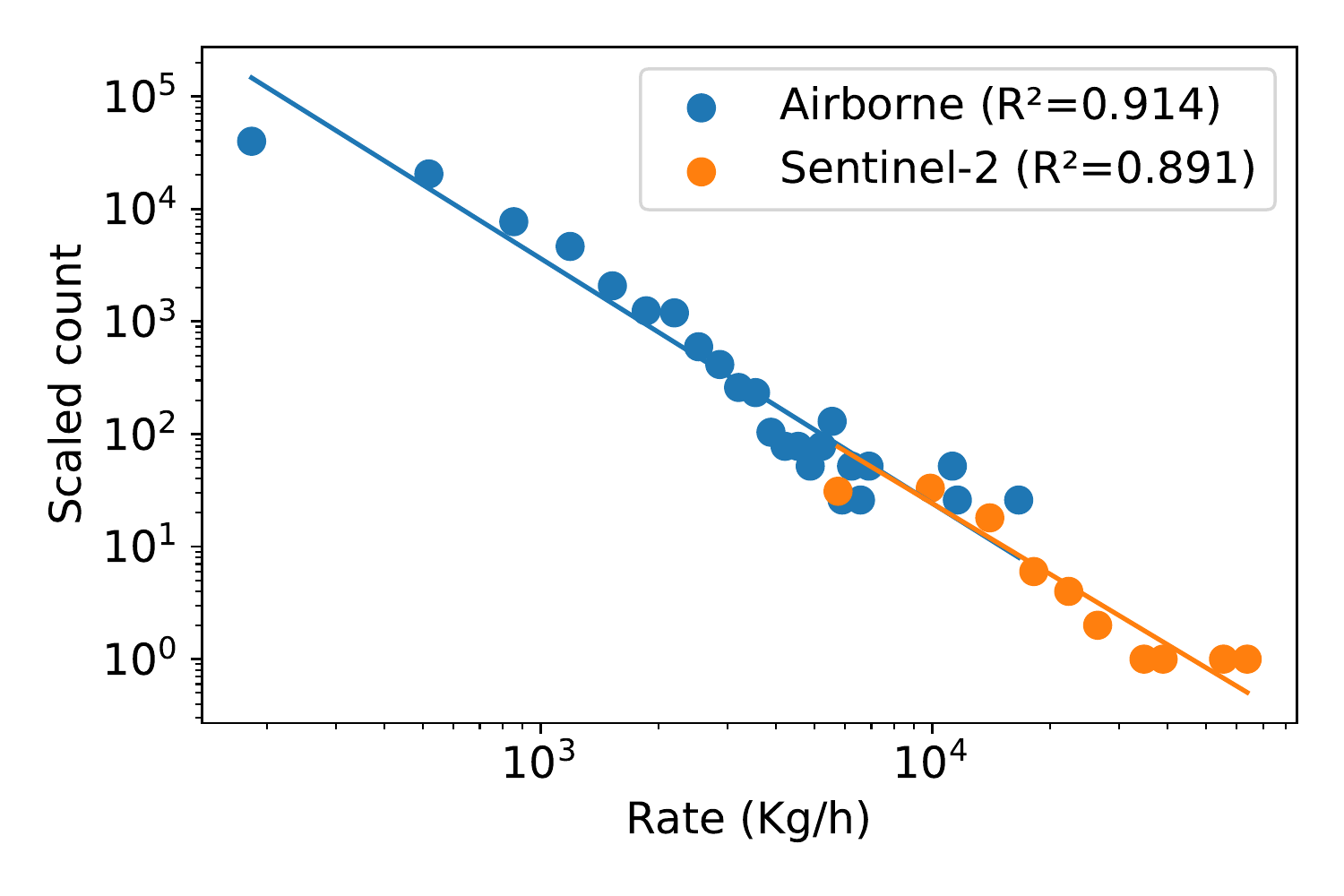}
    \end{subfigure}
    \caption{Local power law for the Permian basin in the US using \emph{Sentinel-2} and airborne~\cite{cusworth2021intermittency} measurements. This seems to validate the hypothesis that \emph{Sentinel-2} is also a proxy indicator for smaller events corresponding to the ones detected by the airborne campaign.}
    \label{fig:powerlaw-permian}
\end{figure}

We performed the same analysis with \emph{Sentinel-2} and airborne measurements in the Permian bassin basin in the US. This power law is presented in Fig.~\ref{fig:powerlaw-permian}. Once again, the power law model seems to be valid at this local regional level. It seems to indicate that the hypothesis such that \emph{Sentinel-2} could be a good proxy indicator for airborne measurements done for the global law power is indeed valid.

\section{Supporting Information}

Plume list with localization and quantification (XLSX)

\begin{acknowledgement}

The authors thank Omar Dhobb from Kayrros for his encouragement and support and acknowledge Edouard Machover for his early work on this subject.
Work partly financed by Office  of Naval research grant N00014-17-1-2552, DGA Astrid project  {``Filmer la Terre"} n\textsuperscript{o} ANR-17-ASTR-0013-01, and MENRT. T. Lauvaux was supported by the French research program {``Make Our Planet Great Again"} (CNRS, project CIUDAD). Alexandre d'Aspremont is at CNRS \& d\'epartement d'informatique, \'Ecole normale sup\'erieure, UMR CNRS 8548, 45 rue d'Ulm 75005 Paris, France,  INRIA  and  PSL  Research  University, and would like to acknowledge support from the {\em ML and Optimisation} joint research initiative with the {\em fonds AXA pour la recherche} and Kamet Ventures, a Google focused award, as well as funding by the French government under management of Agence Nationale de la Recherche as part of the "Investissements d'avenir" program, reference ANR-19-P3IA-0001 (PRAIRIE 3IA Institute).

\end{acknowledgement}

\bibliography{bib}

\end{document}